%% file: main.tex
\documentclass[letterpaper]{article}

\usepackage{prepr}

\usepackage[table]{xcolor}
\usepackage{lipsum}
\usepackage{tcolorbox}
\usepackage{hyperref}
\usepackage{censor}
\usepackage{graphics}
\usepackage{subcaption}

\title{A Survey on Feedback Types in Automated Programming Assessment Systems}
\author{
\PREPauthor{Eduard Frankford}{University of Innsbruck, Department of Computer Science, Innsbruck, Austria}{\\eduard.frankford@uibk.ac.at}
\PREPauthor{Tobias Antensteiner}{University of Innsbruck, Department of Computer Science, Innsbruck, Austria}{\\tobias.antensteiner@uibk.ac.at}
\PREPauthor{Michael Vierhauser}{University of Innsbruck, Department of Computer Science, Innsbruck, Austria}{\\michael.vierhauser@uibk.ac.at}
\PREPauthor{Clemens Sauerwein}{University of Innsbruck, Department of Computer Science, Innsbruck, Austria}{\\clemens.sauerwein@uibk.ac.at}
\PREPauthor{Vivien Wallner}{University of Salzburg, Department of Artificial Intelligence and Human Interfaces, Salzburg, Austria}{\\vivien.wallner@plus.ac.at}
\PREPauthor{Iris Groher}{Johannes Kepler University Linz, Institute of Business Informatics - Software Engineering, Linz, Austria}{iris.groher@jku.at}
\PREPauthor{Reinhold Plösch}{Johannes Kepler University Linz, Institute of Business Informatics - Software Engineering, Linz, Austria}{reinhold.ploesch@jku.at}
\PREPauthor{Ruth Breu}{University of Innsbruck, Department of Computer Science, Innsbruck, Austria}{ruth.breu@uibk.ac.at}
\vspace{0.6cm}
}

\setvenue{the \vspace{3pt}\textbf{ACM Transactions on Computing Education}}
\setvenueone{the \textbf{ACM Transactions on Computing Education}}
\setyear{2025}

\usepackage[numbers]{natbib}

\begin{document}

\maketitle

\newcommand{\revmod}[1]{{\textcolor{black}{#1}}} 

\begin{abstract}
With the recent rapid increase in digitization across all major industries, acquiring programming skills has increased the demand for introductory programming courses. 
This has further resulted in universities integrating programming courses into a wide range of curricula, including not only technical studies but also business and management fields of study.
Consequently, additional resources are needed for teaching, grading, and tutoring students with diverse educational backgrounds and skills. As part of this, Automated Programming Assessment Systems (APASs) have emerged,  providing scalable and high-quality assessment systems with efficient evaluation and instant feedback.
Commonly, APASs heavily rely on predefined unit tests for generating feedback, often limiting the scope and level of detail of feedback that can be provided to students.
With the rise of Large Language Models (LLMs) in recent years, new opportunities have emerged as these technologies can enhance feedback quality and personalization.
To investigate how different feedback mechanisms in APASs are perceived by students, and how effective they are in supporting problem-solving, we have conducted a large-scale study with over 200 students from two different universities. 
Specifically, we compare baseline Compiler Feedback, standard Unit Test Feedback, and advanced LLM-based Feedback regarding perceived quality and impact on student performance.
Results indicate that while students rate unit test feedback as the most helpful, AI-generated feedback leads to significantly better performances. These findings suggest combining unit tests and AI-driven guidance to optimize automated feedback mechanisms and improve learning outcomes in programming education.

\end{abstract}



\maketitle

\input{contents}

\balance

\bibliographystyle{abbrv}
\bibliography{refs}

\end{document}

%% file: contents.tex
\section{Introduction}
\label{sec:introduction}

With digitalization becoming a major topic, many industries are increasingly adopting digital solutions at a staggering pace. Therefore, the need for fundamental programming skills has increased considerably. Programming enables individuals to automate processes, analyze large datasets, and develop custom applications, making it an essential skill for many tasks. This increasing need has also been reflected in many university curricula, where courses such as computational thinking, introductory algorithm design, and basic programming classes are now included in a wider range of programs~\cite{noshin2018teaching,vial2018teaching,hou2020computational,farah2022}.
As a result, the number of students enrolling in programming classes and lectures has grown significantly, creating a pressing demand for high-quality assessment and feedback to support their learning and improve their programming skills~\cite{krusche2020interactive}. However, the time frame for assessments to be evaluated and graded is relatively short, resulting in the need to grade several dozen or even hundreds of assessments per week, depending on the course size. 
Furthermore, solely relying on feedback and assessment from human graders poses the threat of erroneous or inconsistent grading. Additionally,  feedback quality might vary, particularly when grading is performed by tutors or student teaching assistants \revmod{~\cite{henderson2004grading,  bloxham2016let}}.

In this context, APASs have become an essential part of computer science education, offering scalable and efficient means to evaluate programming assignments and provide immediate feedback to students~\cite{krusche2018artemis,Mekterovic2020Building}.
Typically, APASs generate feedback for students by executing pre-defined unit tests on the students' code and returning the output as feedback to the students. However, with the introduction and widespread availability of LLMs, new possibilities for speeding up grading, enhancing feedback quality, and personalization have emerged~\cite{frankford2024ai,bassner2024iris}.  

As of today, only a few empirical studies systematically analyze different types of automated feedback for introductory programming education. Comparisons among baseline Compiler Feedback, standard Unit Test Feedback, and next-generation LLM-based feedback remain largely unexplored, particularly regarding perceived quality and effects on student performance and engagement.
To fill this gap, we have conducted a large-scale survey along with an experiment with first-year bachelor students in introductory programming courses at two different universities.

The main goal of these experiments was to answer the following two research questions:\\

\begin{itemize}[itemsep=0pt, topsep=0pt]
    \item \textbf{RQ1}: How do students perceive different feedback forms when solving programming tasks?   
    \item \textbf{RQ2}: How do different forms of feedback in APASs affect students' success in solving tasks?\\
\end{itemize}

For this purpose, we conducted (1) a qualitative analysis, soliciting student feedback as part of a multipart survey, and (2) a quantitative analysis, where we investigated the students' performance after working on exercises with the different feedback types.
The findings reveal significant differences in how students perceive feedback and perform under different feedback conditions. The survey indicates varying levels of satisfaction and perceived usefulness among the feedback types. The success rates of the students and the quality of their solutions also varied, offering insights into the effectiveness of different feedback mechanisms.

The remainder of this paper is organized as follows. Section~\ref{sec:related_work} reviews related work on feedback types in programming education and the application of AI and LLMs for feedback generation. Section~\ref{sec:applied_research_methodology} describes the applied research methodology, including survey design, exercises, feedback types, data collection methods, data cleaning, and data analysis. Section~\ref{sec:demographics} presents the demographics of the students surveyed. Section~\ref{sec:part1} presents the results regarding the students' individual perceptions of the feedback types, followed by Section~\ref{sec:part2}, which presents the results of the impact of each type of feedback on student performance. Section~\ref{sec:discussion} discusses the implications of the findings, and Section~\ref{sec:limitations} acknowledges the limitations of the study. Finally, Section~\ref{sec:conclusion} concludes the paper and suggests directions for future research.

\section{Background \& Related Work}
\label{sec:related_work}

In this section, we provide a brief introduction to the area of automated assessment tools and programming education and discuss related work in three main areas. First, research on feedback types in automated assessment systems discussing general research on different types of feedback. This includes, for example, more traditional rule-based techniques and new adaptive strategies using AI methods and LLM feedback generation. 
In the second part, we focus on different types of feedback, and how these can affect student performance and can be used to create personalized and meaningful insights into students' programming processes.
Finally, we discuss related studies that have evaluated feedback approaches. Together, these subsections provide a comprehensive analysis of the current state of the art, highlighting the existing gap and potential of current feedback mechanisms in supporting student learning in programming education.
\revmod{Note that this section focuses on a representative subset of feedback types, particularly those commonly used in programming education, and does not aim to cover the full range of automated feedback forms, such as delayed or purely positive feedback.}

\subsection{Automated Assessment Tools \& Feedback Types}
Automated programming assessment tools are essential in computer science education, offering scalable and efficient evaluation of programming assignments. According to Paiva and Leal~\cite{paiva2022}, these systems play a vital role in fostering practical programming skills by analyzing complex program features, supporting diverse task formats, ensuring security, and providing targeted feedback~\cite{paiva2022}. Systems like Edgar~\cite{Mekterovic2020Building}, Artemis~\cite{krusche2018artemis}, Web-CAT~\cite{edwards2008web}, Fitchfork ~\cite{pieterse2013automated}, CodeRunner ~\cite{lobb2016coderunner}, CodeBoard~\cite{truniger2023codeboard}, CodeOcean ~\cite{staubitz2016codeocean}, PythonTutor ~\cite{guo2013online} enable instant feedback to students.

In most cases, the feedback mechanism operates through several key steps. Students first submit their solutions via a Version Control System (VCS)~\cite{ZOLKIFLI2018408}, using either an online code editor provided by the APAS or their local development environment. Once the code is submitted, it triggers a build process on the Continuous Integration (CI)~\cite{6802994} server, which compiles the code and runs predefined unit tests to evaluate the submission. Students can review the test feedback, make corrections, and resubmit their solutions. These systems also support multiple iterations, allowing students to learn from their mistakes and progressively improve their code. Instructors can then monitor students' progress and common issues students face, allowing them to provide targeted assistance and make real-time adjustments to the learning process.

Beyond technical robustness, the effectiveness of feedback generation depends on its impact on student learning behaviors.  Ahmed~\etal~\cite{Ahmed_Srivastava2020} explored the pedagogical benefits of adaptive feedback for novice programmers struggling with compilation errors. While their large-scale randomized study demonstrates improved efficiency in error resolution during lab sessions, it revealed no corresponding improvement in conceptual understanding according to exam performance~\cite{Ahmed_Srivastava2020}. This suggests that automated feedback systems should complement logistical support with strategies that promote deeper learning and conceptual mastery. 

The scalability of feedback generation in large courses adds another layer of complexity, particularly in addressing diverse student errors. 
Singh~\etal~\cite{Singh_Gulwani_Solar-Lezama_2013} introduced a method for automated feedback that utilizes a reference implementation and an error model to derive minimal corrections. A reference implementation is a correct and complete version of a programming solution provided by instructors or system designers. To correct the number of incorrect submissions, it analyzes the student's solution, applies the error model to explore possible corrections, and systematically identifies the minimal set of changes needed to make the solution semantically equivalent to the reference implementation.
However, it was less able to address conceptual errors, highlighting the need for more nuanced feedback mechanisms that address underlying misconceptions~\cite{Singh_Gulwani_Solar-Lezama_2013}. The work of Marin~\etal~\cite{Marin_Pereira_Sridharan_Rivero_2017} demonstrates the use of a semantic-aware technique to enhance feedback personalization in introductory Java Massive Open Online Courses (MOOCs). Using extended program dependency graphs and subgraph matching, their system provided semantically rich, personalized feedback at scale. However, their approach also highlights the complexity of creating feedback that is both individualized and broadly applicable across diverse programming solutions~\cite{Marin_Pereira_Sridharan_Rivero_2017}.

Recent advancements in Machine Learning (ML) and Artificial Intelligence (AI) have enabled the development of systems that provide tailored student feedback, significantly enhancing learning outcomes and engagement. Kochmar~\etal~\cite{Kochmar2020Automated} proposed an ML approach using Natural Language Processing (NLP) to deliver personalized hints and explanations based on individual needs, demonstrating improvements in both learning and students’ perceptions of feedback quality~\cite{Kochmar2020Automated}. Similarly, Ahmed~\etal~\cite{ahmed2020} introduced TEGCER, an AI-powered tool for novice programmers that employs supervised classification to match code submissions with previously observed error patterns, leveraging a dataset of over 15,000 error-repair examples~\cite{ahmed2020}. TEGCER provides tailored solutions to specific errors, enabling faster debugging. In a study with over 230 participants, students using TEGCER resolved errors more efficiently than those relying on human tutors, underscoring the potential of AI-driven tools to enhance programming education through personalized and efficient support.

Building on these advancements, Frankford~\etal~\cite{frankford2024ai} utilized LLMs to generate context-aware feedback, demonstrating their capability to deliver instantaneous, personalized support at scale~\cite{frankford2024ai}. Bassner~\etal~\cite{bassner2024iris} further validated the effectiveness of LLMs in enhancing the learning experience by tailoring feedback to individual student submissions~\cite{bassner2024iris}. 

These findings highlight the potential of LLMs as powerful tools in modern, scalable educational systems. Complementing this, Barrow~\etal~\cite{Barrow2008Assessing} investigated feedback mechanisms in Intelligent Tutoring Systems (ITSs), focusing on the inclusion of positive reinforcement alongside traditional negative feedback. Their study found that students receiving both positive and negative feedback solved problems more quickly and with fewer attempts, all while mastering the same number of concepts as those in a control group receiving only negative feedback~\cite{Barrow2008Assessing}. These results emphasize the importance of balanced feedback strategies in AI-driven educational tools, showcasing how a combination of reinforcement types can enhance learning efficiency and overall performance.

\subsection{Effects of Different Feedback Types on Students' Performance}

Several studies have shown that immediate and detailed feedback can significantly improve learning outcomes of students~\cite{Zampirolli2021An, Aleman2011Automated, Kleij2013Effects}. Paiva~\etal~\cite{paiva2022} conducted a systematic review on automated feedback in programming environments, highlighting the importance of timely, specific, and actionable feedback~\cite{paiva2022}. Ulla~\etal~\cite{ullah2018} mentioned that automated assessment systems have an impact on novice programmers. This impact is crucial for students in introductory programming courses as it provides immediate feedback and eases the teacher's workload when the number of students is high, so that the development of essential skills can be supported~\cite{ullah2018}. Cavalcanti~\etal~\cite{cavalcanti2021automatic} mentioned that feedback is crucial for learning, especially in online courses where instructors and students are physically separated. Automatic feedback systems have been proposed to address this, and a systematic review~\cite{cavalcanti2021automatic} reveals that automatic feedback increases student performance.  Keuning~\etal~\cite{keuning2018systematic} compared how feedback is generated and found that the most common technique is automated testing using unit tests, which is used by 58.4\% of the tools, followed by program transformation techniques (37.6\%) and static analysis (36.6\%)~\cite{keuning2018systematic}.

Chen~\etal~\cite{chen2020analysis} conducted an experiment involving 76 computer science (CS) students to investigate the impact of different feedback designs on student learning and interaction. They examined three groups, each receiving one of the following feedback categories: (1) Knowledge of Results (KR): includes only information on which test cases failed; (2) Knowledge of Correct Responses (KCR): includes KR plus additional information on where the failure occurred within the test case; (3) Elaborated Feedback (EF): includes KCR plus a hint on what the student might need to do to fix their code. Across three complex programming assignments, they found that students who received higher levels of feedback (KCR and EF) significantly outperformed those who received only KR feedback. The EF feedback was implemented as a one-level hint addressing the top five mistakes from prior student submissions, with additional explanations provided~\cite{chen2020analysis}. \revmod{The study, comparing three different feedback types, shows that more detailed feedback significantly improves student performance.}

Qian and Lehman~\cite{Qian2019UsingTF} investigated the impact of targeted feedback on addressing misconceptions in high school students' Java programming. Their study demonstrated that tailored feedback, designed to address specific errors and misunderstandings, not only helped students correct their mistakes but also reduced the occurrence of intermediate errors in subsequent attempts. By offering explanations of reasons for errors and guiding students toward proper problem-solving strategies, the feedback promoted deeper conceptual understanding. These findings emphasize the value of targeted feedback in improving learning outcomes and align with the growing interest in AI-driven feedback generation, which offers the potential to deliver personalized, context-aware support at scale~\cite{Qian2019UsingTF}. \revmod{The feedback used here can be categorized as targeted and corrective, aimed specifically at conceptual misconceptions in a high school programming context. Its effectiveness was reflected in reduced error rates and deeper learning.}

Similarly, Clegg~\etal~\cite{Clegg_McMinn_Fraser_2020} highlighted how characteristics of test suites, such as coverage, size, and redundancy, can significantly impact grading outcomes. Their analysis focused on optimizing test suites to achieve comprehensive code coverage while minimizing redundancy, ensuring consistent and equitable feedback~\cite{Clegg_McMinn_Fraser_2020}. Such optimization not only enhances the accuracy of automated grading systems but also fosters student trust by maintaining grading integrity. \revmod{This study highlights how the design of assessment infrastructure affects the reliability and fairness of outcome-oriented feedback in grading contexts.}

Assessment systems have also proven effective in supporting continuous improvement and encouraging timely assignment completion. For example, Yan~\etal~\cite{Yan_Wu_Nguyen_Chen_2020} evaluated the ProgEdu system and found that its iterative feedback mechanisms significantly enhanced student learning outcomes and code quality~\cite{Yan_Wu_Nguyen_Chen_2020}. However, their study also identified areas for improvement, such as enhancing feedback precision and usability, highlighting the need for ongoing refinement to meet evolving learner needs. \revmod{The feedback provided in this study can be classified as process-oriented, helping students improve their programming skills with repeated submissions. }

Comparative analyses of automated grading tools illustrate the trade-offs between different assessment methodologies. Bey~\etal~\cite{Bey_Jermann_Dillenbourg_2017} compared Algo+, a static analysis-based system, with the EPFL grader, which relies on unit tests~\cite{Bey_Jermann_Dillenbourg_2017}. While Algo+ excelled at identifying partially correct solutions, it sometimes underestimated novel correct submissions absent from its reference set. This finding emphasizes the importance of aligning grading methodologies with specific educational objectives to maximize their pedagogical impact.

\subsection{Methodologies Used to Evaluate Feedback Forms}
Understanding the effectiveness of feedback in APASs is essential for improving learning outcomes, fostering motivation, and supporting skill development. Research in this area has explored various approaches to feedback delivery, analyzing their impact on students’ educational experiences and instructors' workflows.

One widely adopted approach for evaluating feedback mechanisms in APASs is through student surveys. Messer~\etal~\cite{messer2024} conducted an extensive review of 121 papers published between 2017 and 2021, focusing on evaluation methodologies used in automated grading tools for programming education. Their findings highlighted the frequent use of student surveys to assess aspects such as satisfaction with instant feedback, the clarity and usefulness of error messages, and the opportunities for iterative learning enabled by resubmissions. These surveys typically include Likert-scale questions, open-ended prompts, and comparative assessments in which students evaluate automated tools against traditional teaching methods. The results consistently show that students value the immediacy and consistency of automated feedback, particularly its effectiveness in identifying specific coding issues and providing actionable suggestions for improvement. These findings underscore the critical role of feedback beyond correctness, emphasizing its importance in supporting formative learning processes that foster reflection and improvement~\cite{messer2024}.

The integration of gamification into APASs introduces an innovative dimension to feedback evaluation. Polito and Temperini~\cite{polito2021} developed a gamified learning platform that combines immediate feedback with motivational features such as experience points, badges, and leaderboards to enhance engagement and skill development. Their experimental study demonstrated that real-time feedback, combined with gamification, significantly boosts student engagement~\cite{polito2021}. A detailed post-experiment survey of 36 questions and a 5-point Likert scale assessed students' experiences, engagement levels, and perceptions of the platform's usefulness in skill development. The survey also explored the impact of gamification elements and students’ willingness to adopt similar systems in other educational contexts. The results were overwhelmingly positive, with students rating their overall experience highly with 4.19 out of 5. Features like immediate feedback and personalized leaderboards were particularly appreciated for their motivational impact. These findings align with improved academic performance, showcasing the potential of gamified APASs to facilitate iterative learning and promote deeper student engagement in programming education.

Another example analyzed by \revmod{J. English and T. English} is the Checkpoint system, which provides detailed, iterative feedback with unlimited submission attempts ~\cite{english2015}.
A survey of 141 students, incorporating 15 Likert-scale questions and open-ended prompts, revealed a strong appreciation for Checkpoint's usability, reliability, and support for self-paced learning. Students particularly valued the opportunity to make multiple attempts, as it helped them identify and correct mistakes through detailed, actionable feedback. Faculty also reported reduced grading workloads and access to granular performance analytics for targeted interventions. However, some students expressed frustrations with the rigidity of the feedback and suggested greater flexibility to accommodate diverse learning styles. These findings highlight the strengths of automated systems like Checkpoint in enhancing the learning process while pointing to the need for balancing automation with adaptability~\cite{english2015}.

Substantial research compares various feedback mechanisms with traditional non-APAS-based approaches \cite{keuning2018systematic, Clegg_McMinn_Fraser_2020}, yet the differences in student performance and perceived usability for compiler feedback, unit-test feedback, and AI-based feedback in APASs remain unexplored. Notably, no large-scale, controlled study has compared these feedback types with a representative sample. This research addresses that gap by systematically evaluating three feedback conditions: (1) no support as a baseline with compiler feedback only, then (2) unit-test feedback, and lastly, (3) AI-generated feedback, and examining their impacts on student performance, usability, intention to use, and output quality. To achieve this, it applies diverse survey methodologies \cite{messer2024, polito2021, english2015, ihantola2010, alamutka2005} that capture students’ perceptions and responses, offering valuable insights into each feedback type's effectiveness.

While several authors have investigated different types of feedback and their impact on student performance, there is still a lack of research on assessing and comparing these types of feedback, particularly regarding their evaluation through student surveys with representative sample sizes. So far, only a few studies have systematically analyzed how these feedback types differ concerning ease of use, perceived usefulness, and impact on learning outcomes. Furthermore, the use of LLMs for feedback generation is a novel approach and has thus never been directly investigated in a study comparing all three feedback types. Therefore, our work presented in this article helps to create a comprehensive overview to better understand, compare, and optimize feedback mechanisms in programming education.


\section{Research Methodology}
\label{sec:applied_research_methodology}

To systematically investigate the use of different forms of feedback in APASs for introductory programming education, we adopted the Goal Question Metric (GQM) approach to define our research goal and questions~\cite{van2002goal}. The GQM framework provided a structured means to evaluate the perceptions and effectiveness of various feedback types on students’ learning outcomes.  The overall goal, the derived research questions, and constituent metrics are outlined in \citefig{gqm}.

\begin{figure*}[h!]
\centering\includegraphics[width=0.95\textwidth]{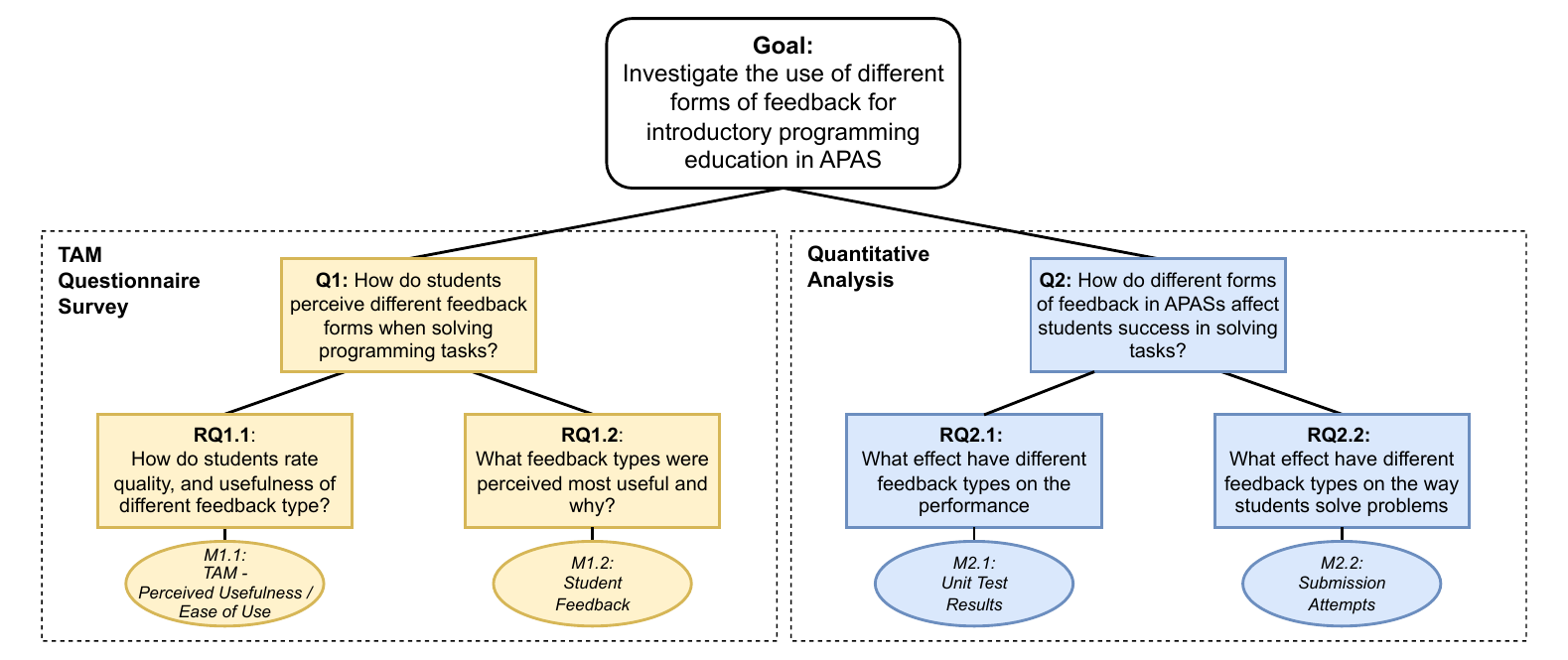}
    \caption{Goal Question Metric (GQM) Framework for Investigating Feedback Types in APASs}
    \label{fig:gqm}
    \vspace{-0.55em}
\end{figure*}

The study's overall goal, as illustrated in \citefig{gqm}, was to explore the effects of feedback types on both student perceptions and their success in solving programming tasks. To achieve this, two primary research questions (RQs) were developed, each with specific sub-questions:

\begin{itemize}
    \item \textbf{RQ1: How do students perceive different feedback forms when solving programming tasks?}
        \begin{itemize}
            \item \textbf{RQ1.1:} How do students rate the quality and usefulness of the different feedback types?
            \item \textbf{RQ1.2:} What feedback types were perceived as most useful and why?
        \end{itemize}
    \item \textbf{RQ2: How do different forms of feedback in APASs affect students’ success in solving tasks?}
        \begin{itemize}
            \item \textbf{RQ2.1:} What effect do different feedback types have on performance?
            \item \textbf{RQ2.2:} What effect do different feedback types have on the way students solve problems?
        \end{itemize}
\end{itemize}

The study was conducted at two universities in Austria from Autumn 2023 to Spring 2024, integrating the survey into students’ coursework to capture a broad set of student responses. The exercises and experimental environment were implemented in the online editor provided by the APAS \textit{Artemis}\footnote{\url{https://github.com/ls1intum/Artemis}}, enabling real-time feedback and interaction with the platform. \revmod{The feedback was generally triggered by students clicking on the submit button in Artemis. The students could click on the submit button as many times as they want.}

The research methodology combined a \textit{Technology Acceptance Model (TAM) questionnaire-based survey} with \textit{quantitative analysis of student performance}:

\begin{itemize}[itemsep=0pt, topsep=0pt]
    \item \textbf{Survey Integration (RQ1):}
    Students answered TAM-based surveys immediately after solving each programming exercise in Artemis to assess their perceptions of feedback quality, usefulness, and ease of use. A final open-ended questionnaire gathered qualitative insights into their preferences and experiences with the different feedback types.\\

    \item \textbf{Experiment with Three Exercises (RQ2):}
    \revmod{All students experienced all feedback types as they worked on three programming exercises, where each exercise was paired with one of the three feedback types:}
    \begin{itemize}
        \item \textit{Compiler Feedback} (baseline: syntax and runtime errors only),
        \item \textit{Unit Test Feedback} (structured correctness verification),
        \item \textit{AI-generated Feedback} (adaptive, hint-based guidance).
    \end{itemize}
    All code submissions, time between attempts, and final correctness scores were logged for detailed analysis.\\

    \item \textbf{Data Collection and Metrics:}
    \begin{itemize}
        \item \textit{M1.1:} TAM constructs perceived usefulness (PU), ease of use (EU) were rated using Likert-scale questions to measure subjective feedback quality.
        \item \textit{M1.2:} Post-experiment feedback was qualitatively analyzed to understand why students favored certain feedback types.
        \item \textit{M2.1:} Unit test results were used to evaluate the correctness of students' final solutions under different feedback conditions.
        \item \textit{M2.2:} Submission logs provided insights into problem-solving strategies, including iteration behavior and time spent refining solutions.
    \end{itemize}
\end{itemize}

The study ensured diverse representation by including students from two universities, enabling broader generalizability of findings. This methodological approach provided a comprehensive evaluation of both subjective and objective impacts of feedback types in APASs. Survey materials, including exercises, tests, and questionnaires, are described in more detail in the following subsections.

\subsection{Exercise Design and Selection}

To ensure a valid comparison across feedback types, we designed a set of exercises that were consistent in size and complexity. As the experiments were conducted in introductory Java courses at both universities in the first half of the semester, we selected the topic of Arrays, which aligned with both course curricula.

Specifically, we selected examples covering aspects of one-dimensional and two-dimensional arrays, as this provided sufficient complexity to assess competencies in this area compared to earlier lectures covering topics such as operators, variables, loops, or simple conditions. As the topic of arrays had been covered in the first half of the semester in both universities, all students had similar prior experience with arrays through previous assignments.
Each exercise was developed to target similar competencies based on an established competence model for introductory Java courses~\cite{Ploesch2024}. The initial drafts were created by one instructor and then reviewed and refined by several faculty members from both universities who were involved in the introductory programming courses. This iterative refinement step ensured the consistency and similar levels of complexity of the exercises.  In the following, we provide a brief description of the three exercises included in the study and the respective development steps.


\subsubsection{Exercise 1: Mars Rover}

This exercise involves helping a Mars Rover navigate a defined zone. The Mars surface is represented as a matrix, where \texttt{'*'} indicates a free space and \texttt{'R'} marks the rover's current position. The task requires students to write a method \texttt{'moveRover'} that moves the rover within the matrix based on given row and column values. If the movement takes the rover outside the matrix, it stops at the edge. The method updates the matrix by changing the start position from \texttt{'R'} to \texttt{'*'} and the end position from \texttt{'*'} to \texttt{'R'}, and then returns the updated matrix. If the matrix is null or empty, the method returns null. Positive row values move the rover right, negative row values move it left. Positive column values move it down, and negative column values move it up. The exercise provides the helper methods,\texttt{'searchRover'} and \texttt{'displayMapFromMatrix'}, for finding the rover's position and testing output. 

\subsubsection{Exercise 2: Array Transformation}

In this exercise, students write a method \texttt{'calc4All'}  that applies an arithmetic operation to each element in an integer array using a specified operand, storing the results in a new array. If the input array is empty or null, or if the operation is invalid (`+,` `-,` `*,` `/,` and `\%` are the valid operators), the method returns null. Special cases where the operand cannot be zero are handled by returning null. Students use the provided \texttt{'arrayToString'} method to test their output. 

\subsubsection{Exercise 3: Matrix Validation}

This exercise requires students to write a method  \texttt{'isValidNumberMatrix'} to verify if a 3x3 matrix contains the numbers 1-9 without duplicates. The matrix should not contain other numbers, and if it fails these criteria or is null/not 3x3, the method returns false. This task reinforces students' understanding of array handling and validates their matrix manipulation skills. 

\subsubsection{Exercise Development Process}

Each exercise was developed using the same structured process to ensure comparable difficulty and alignment with regard to the needed competencies to solve the exercises. For the development of the exercises, we followed the process described by Plösch et al.~\cite{Ploesch2024}. The stages were as follows:

\begin{enumerate}[itemsep=0pt, topsep=0pt]    
\item \textbf{Initial Draft and Solution Development}: An instructor selected the competencies from the competence model and prepared initial drafts of the exercises, including task descriptions and sample solutions.
    \item \textbf{First Review}: Two researchers reviewed each exercise, provided feedback, and independently implemented solutions. Based on this feedback, the exercises were revised for consistency in complexity and code length. The researchers also reviewed the mapping of the competencies to the exercises and sample solutions. 
    \item \textbf{Second Review by Faculty}: Faculty from both universities reviewed the mapped competencies and verified the appropriateness of the exercises. Minor adjustments were made, such as adding required checks for boundary and invalid inputs, to ensure alignment across exercises.
    \item \textbf{External Review and Final Adjustments}: Two external instructors, experienced in teaching introductory Java courses but not involved in the initial creation, conducted an independent review. They suggested minor changes to wording and formatting, which were incorporated to improve clarity.
\end{enumerate}

Each finalized exercise provided students with a clear problem description, along with example input and expected output. Exercises 1 and 3 involved two-dimensional arrays, while Exercise 2 focused on a one-dimensional array with additional operations. Our collaborative review process helped ensure that each exercise was balanced in terms of difficulty, enabling a reliable assessment of feedback types across institutions.

After having developed well-reviewed exercise specifications including a mapping to competencies, we started to design and implement competency-based unit tests. Again, we followed the process proposed in~\cite{Ploesch2024}. For each assigned competence, we implemented at least one unit test. For complex competencies (e.g., iterating over two-dimensional arrays) more unit tests were necessary. We added further tests necessary to check the correctness of the solution. Coverage analysis of the sample solution helped us to identify missing test cases.


\subsection{Study Execution and Data Collection Process}

At both universities, the study was carried out in the first half of the semester as part of a regular 90-minute in-person class session.  As illustrated in \citefig{overview}, the study was designed to systematically collect survey and performance data from students at multiple stages, gathering both quantitative and qualitative insights.
This structured data collection process allowed for an in-depth evaluation of student experiences with each feedback type.
\begin{figure*}[h!]
    \centering
\includegraphics[width=0.8\linewidth]{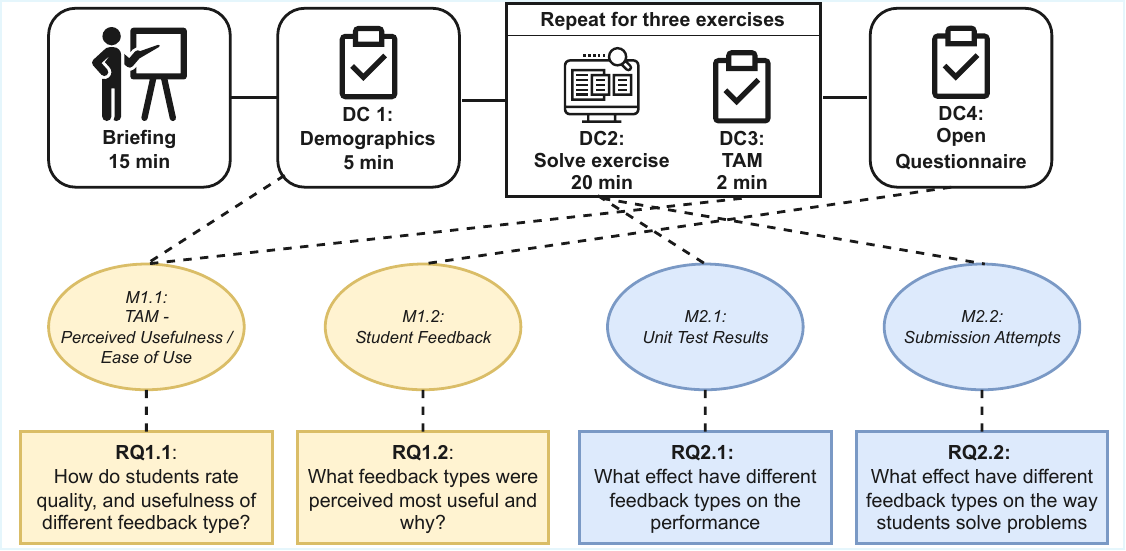}
    \caption{Overview Study Workflow}
 
    \label{fig:overview}
\end{figure*}
The 90-minute session was split into three parts, starting with a 15-minute introduction describing the structure of the survey and exercise, and asking for consent. 
The study contained four key data collection points \revmod{(in Fig. \ref{fig:overview} defined as DC 1-4) (three during the 90-minute session (DC1-3), and one (DC4)} post-study questionnaire distributed after the session to give the students additional time to reflect on the exercises and feedback mechanisms). Each data collection point targeted specific aspects of the study:

\subsubsection{Data Collection Point 1: Demographics Questionnaire}
\label{DC1}

After the initial briefing, before starting the exercises, students completed a 5-minute demographics questionnaire to gather baseline information, which included the following questions:

\begin{itemize}[itemsep=0pt, topsep=0pt]
    \item \textbf{Gender:} Options included Male, Female, Diverse/Other, Prefer not to say.
    \item \textbf{Age:} Students specified their age in years.
    \item \textbf{Study Program Enrollment:} Various Bachelor's and Master's programs were listed, with an option to specify other programs.
    \item \textbf{Current Semester:} Options ranged from 1st to 6th semester, with an option to specify other semesters.
    \item \textbf{Programming Languages Used Before This Course:} Multiple options were provided for students to indicate prior experience with specific programming languages, along with a field for additional languages.
    \item \textbf{Years of Programming Experience:} Students specified their years of experience.
    \item \textbf{Highest Educational Qualification:} Options included different levels from high school to doctorates, with an option to specify additional qualifications.
    \item \textbf{Number of Times Taking the Course:} Students indicated how many times they had enrolled in the course.
    \item \textbf{Experience with Artemis:} Students were asked if they had previously used the Automated Programming Assessment System (Yes/No).
\end{itemize}

\noindent
Additionally, students were prompted to create a personal code, to enable anonymous tracking across all survey responses. This was to ensure that only valid responses were included in the subsequent analysis of results.

\subsubsection{Data Collection Point 2: System Logs and Exercise Submission Data}

As students worked on the programming exercises, each submission to the APAS (Artemis) was logged with the following data. Students were permitted to submit any number of ``intermediary'' solutions and received feedback for each version of the (partial) submission.

\begin{itemize}[itemsep=0pt, topsep=0pt]
    \item \textbf{Code Submitted:} A snapshot of the student's solution at the time of each submission.
    \item \textbf{Feedback Provided:} The type of feedback the student received for each submission.
    \item \textbf{User Identifier and Timestamp:} Each submission was tagged with a unique user ID and timestamp to track progress and interaction.
\end{itemize}

\noindent
After the study, all final submitted solutions were further evaluated through an additional set of unit tests created by independent instructors, ensuring an unbiased assessment of solution quality.

\subsubsection{Data Collection Point 3: Post-Exercise Feedback Questionnaires}

Following each exercise, students completed a TAM-based questionnaire tailored to the specific feedback type. Each survey included the following questions, with responses on a 7-point Likert scale from 'Strongly Disagree' to 'Strongly Agree':

\begin{itemize}[itemsep=0pt, topsep=0pt]
    \item \textbf{Speed:} "The use of Artemis with [Feedback Type] enables me to complete programming tasks faster."
    \item \textbf{Simplification:} "The use of Artemis with [Feedback Type] makes it easier for me to accomplish programming tasks."
    \item \textbf{Usefulness:} "The use of Artemis with [Feedback Type] is useful for accomplishing programming tasks."
    \item \textbf{Future Use:} "In the future, I would use Artemis with [Feedback Type] to solve programming tasks."
    \item \textbf{Clarity:} "The interaction with Artemis and [Feedback Type] is clear and understandable."
    \item \textbf{Effort:} "The interaction with Artemis and [Feedback Type] does not require much mental effort."
    \item \textbf{Ease of Use:} "Artemis with [Feedback Type] is easy to use."
    \item \textbf{Functionality:} "I find it easy to get Artemis with [Feedback Type] to do what I want it to do."
    \item \textbf{Feedback Quality I:} "The quality of the received [Feedback Type] is high."
    \item \textbf{Feedback Quality II:} "I had no problems with the quality of the [Feedback Type]."
\end{itemize}

\noindent Each post-exercise questionnaire took approximately 2-3 minutes, enabling immediate reflection on the feedback type used in each exercise.

\subsubsection{Data Collection Point 4: Post-Experiment Open-Ended Questionnaire}

After completing the main part of the exercises, students were asked to complete a final online survey designed to capture qualitative insights into their overall experiences and preferences and collect additional feedback and comments. The survey included the following items:

\begin{itemize}[itemsep=0pt, topsep=0pt]
    \item \textbf{Personal Code:} Students entered their unique code to connect responses with the demographics questionnaire.
    \item \textbf{General Comments:} "Please provide any general comments about the study, including your thoughts on the feedback types and how they were used."
    \item \textbf{Most Helpful Feedback Type:} Students were asked to select the feedback type they found most helpful from AI Feedback, Compiler Feedback, or Unit Test Feedback.
    \item \textbf{Reason for Feedback Preference:} "Explain why you found the selected feedback type the most helpful. Describe specific aspects or functions you found beneficial."
    \item \textbf{Suggestions for Additional Feedback Types:} "Are there any other feedback types not currently available in Artemis that you would find helpful? If so, please describe them."
\end{itemize}

\noindent This post-experiment questionnaire allowed students to express their preferences, highlight the strengths and weaknesses of each feedback type, and provide suggestions for future feedback enhancements in Artemis.

\subsection{Feedback Types}
\label{subsec:feedback_types}

To assess the effectiveness of various feedback mechanisms, we employed three distinct feedback types and implemented means of support for the programming tasks within the Artemis APAS.

\begin{itemize}
    \item \textbf{No Feedback Support (Compiler Output Only):} This basic level of feedback provides the baseline of support to be measured and includes only standard compiler output, such as syntax errors, warnings, and runtime errors. It lacks additional hints or solutions, thereby simulating a ``minimal support environment``, as students, for example, would also experience in a traditional desktop-based IDE (without any additional extended support plug-ins, such as CoPilot~\cite{copilot}). This feedback type allowed us to investigate students’ problem-solving strategies when relying solely on standard debugging tools.
    \item \textbf{Traditional Unit Test Feedback:} This feedback type employs predefined unit tests (JUnit in our setting) to validate students' code for correctness against specific test cases. Traditional Unit Test Feedback is commonly used in many APASs and offers structured feedback that indicates whether the code meets the functional requirements, identifying any failed test cases. 
    \item \textbf{AI-Generated Feedback:} Using OpenAI's GPT-3.5-Turbo \footnote{\url{https://platform.openai.com/docs/models/gpt-3.5-turbo}} model, this adaptive feedback type acts as a supportive ``tutor``. It provides students with targeted hints and guidance to improve their code, encouraging independent debugging rather than directly revealing the solution.
\end{itemize}

\subsubsection{Compiler Feedback}
Compiler Feedback refers to the default outputs generated by the compiler upon code submission. This includes syntax errors, runtime messages, and any print statements that students may use to debug their code. To simulate Compiler Feedback within our APAS, we configured the system to display program output via a pop-up window (\citefig{compilerfeedback2}). Since Artemis does not support direct code execution without test cases, we implemented a placeholder failing test case to trigger the display of all outputs. Students were informed about this setup, ensuring they understood that they should continue refining their code until it met the exercise requirements.

\begin{figure}[h!]
    \centering
    \includegraphics[width=.99\linewidth]{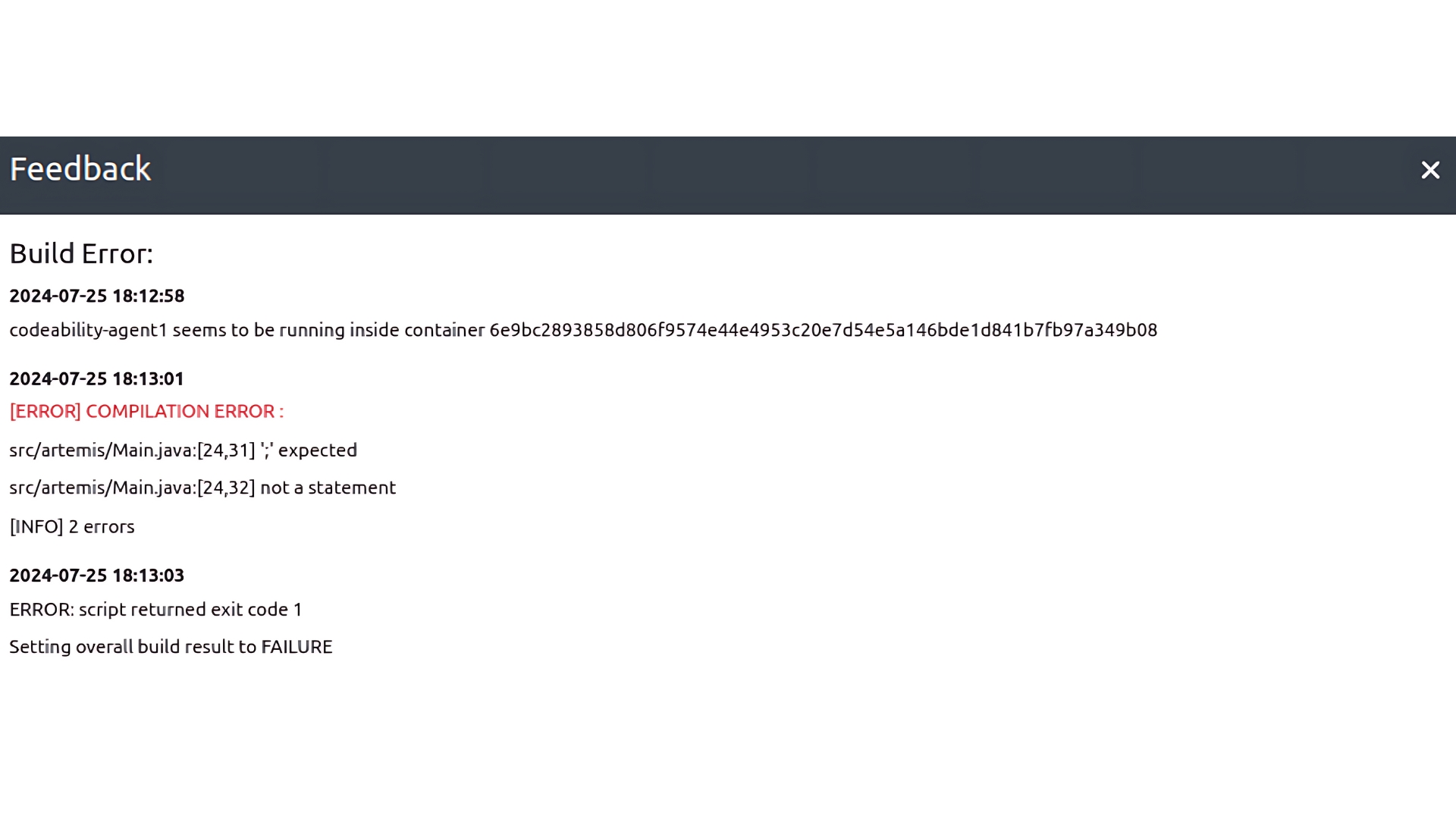}
    \caption{Example of Compile Error Feedback}
    \label{fig:compilerfeedback1}
\end{figure}

\noindent If a submission failed due to a compile error, a build error message was displayed in the pop-up (see Figure~\ref{fig:compilerfeedback1} for an example of a missing semicolon error). This setup allowed students to receive basic feedback without any structured guidance on code correctness.

\begin{figure}[h!]
    \centering
\includegraphics[width=.99\linewidth]{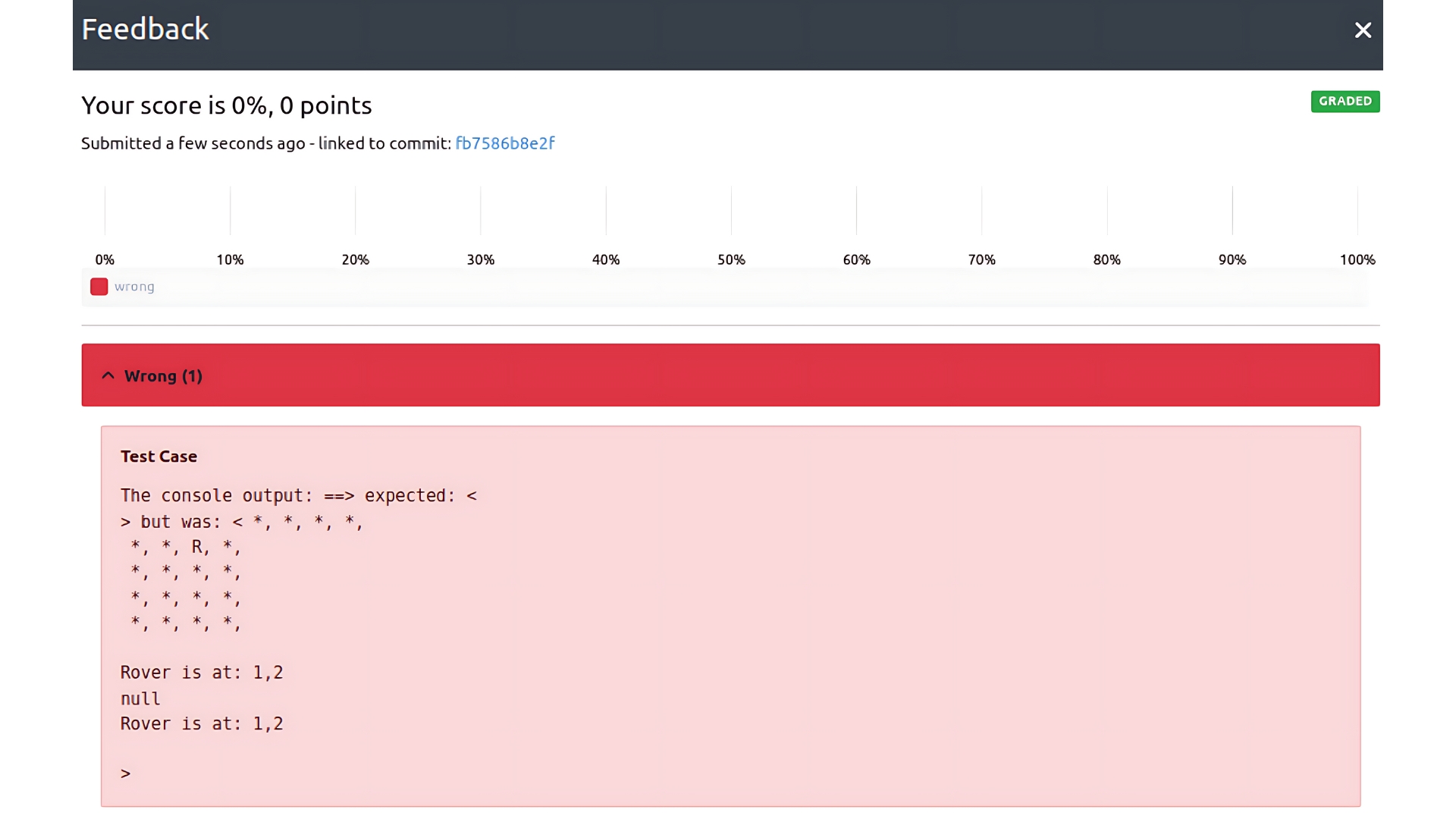}
    \caption{Example of Compiler Feedback Output}
    \label{fig:compilerfeedback2}
\end{figure}

\begin{figure}[h!]
    \centering
    \includegraphics[width=.99\linewidth]{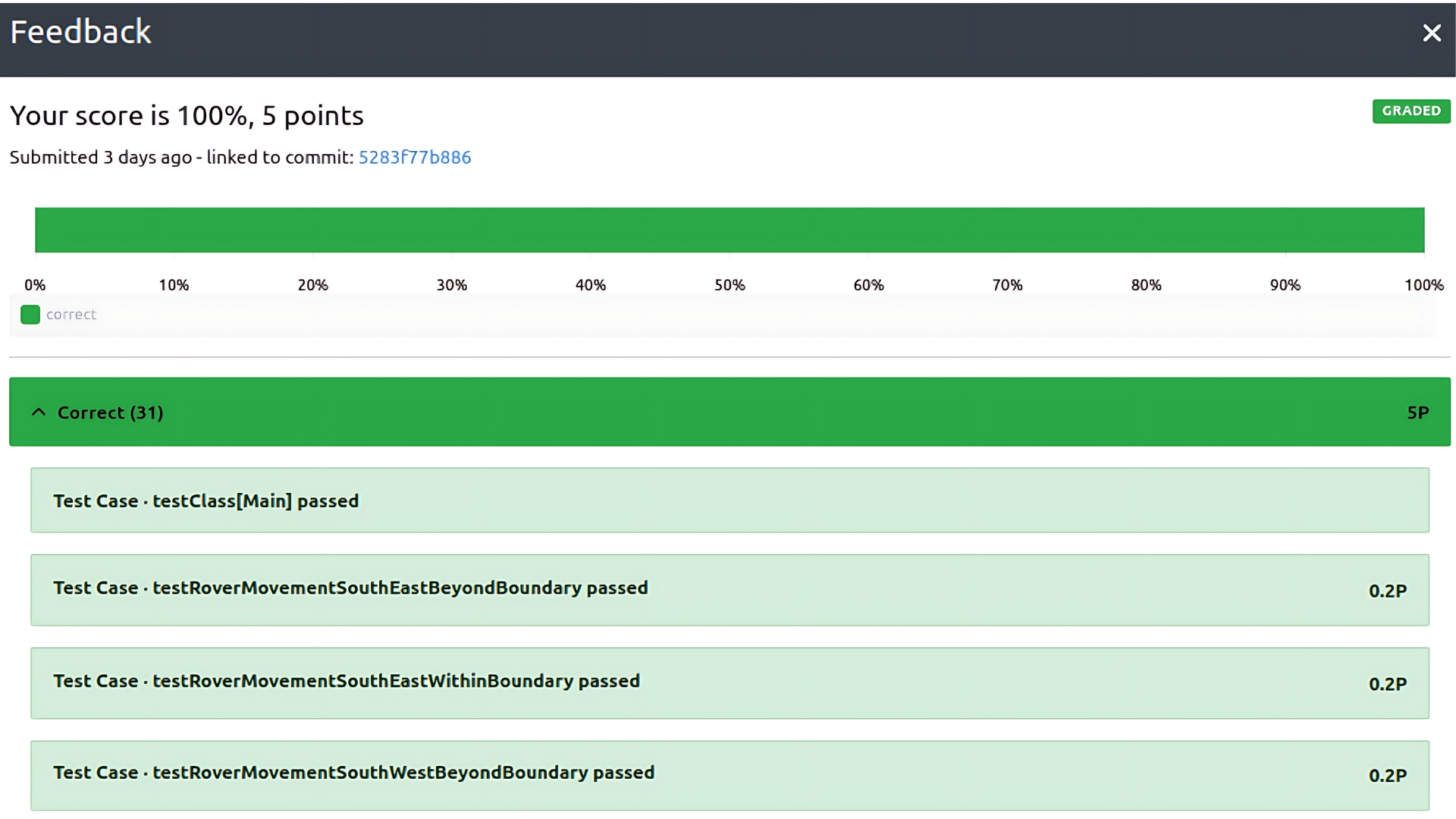}
    \caption{Example of Traditional Unit Test Feedback}
    \label{fig:unittestfeedback1}
\end{figure}

\subsubsection{Unit-Test Feedback}
For traditional feedback, unit tests were used to evaluate the functional correctness of students' code submissions. Upon submission, the system executed a set of predefined unit tests developed by instructors to validate key aspects of the solution. Students received feedback on passed and failed test cases, with detailed error messages indicating specific areas of failure, as shown in \citefig{unittestfeedback1}. This feedback type mirrors the structured, task-oriented feedback used in many APASs, helping students align their implementations with specific functional requirements.

To ensure quality, the unit tests underwent a verification process, similar to the exercise descriptions, by the instructors who authored the exercises (cf. exercise creation process above).

\subsubsection{AI-Generated Feedback}
AI Feedback was generated by the GPT-3.5-Turbo model and designed to provide adaptive, tutor-like guidance. When a student submitted a solution, the server sent the student's solution and the exercise description via an API request to OpenAI's GPT-3.5-Turbo model with the instructions to return hints rather than solutions (see Prompt~\ref{prompt1}). This format allowed targeted feedback that encouraged self-driven debugging and improvement. GPT-4 and newer models, which offer potential improvements in natural language understanding and feedback generation, were not utilized in this study because they were not yet available at the time the study was conducted.\\

\noindent\begin{minipage}{\columnwidth}
\begin{chatgptprompt}[title=Prompt 1, label={prompt1}]
AAAct as a tutor and help the student debug his or her code. Do not provide the complete solution; only give hints and speak to the student directly. If you think the code already fulfills the problem statement, inform the student. The code currently looks like this: \textbf{[Inserted Code]} 

This is the problem the student is trying to solve: \textbf{[Inserted Problem Statement]}
\end{chatgptprompt}
\captionof{prompt}{GPT-3.5-Turbo Prompt for AI-Generated Feedback}
\vspace{1em}
\end{minipage}

The AI model's responses were displayed within the Artemis interface (\citefig{aifeedback}), helping students identify issues without offering explicit solutions. The feedback request was initiated via an API call, executed within a failing test case, as the Artemis system requires test cases to display outputs. Students were briefed to disregard the failing status and focus on the AI's feedback hints.

\begin{figure}[H]
    \centering
    \includegraphics[width=.99\linewidth]{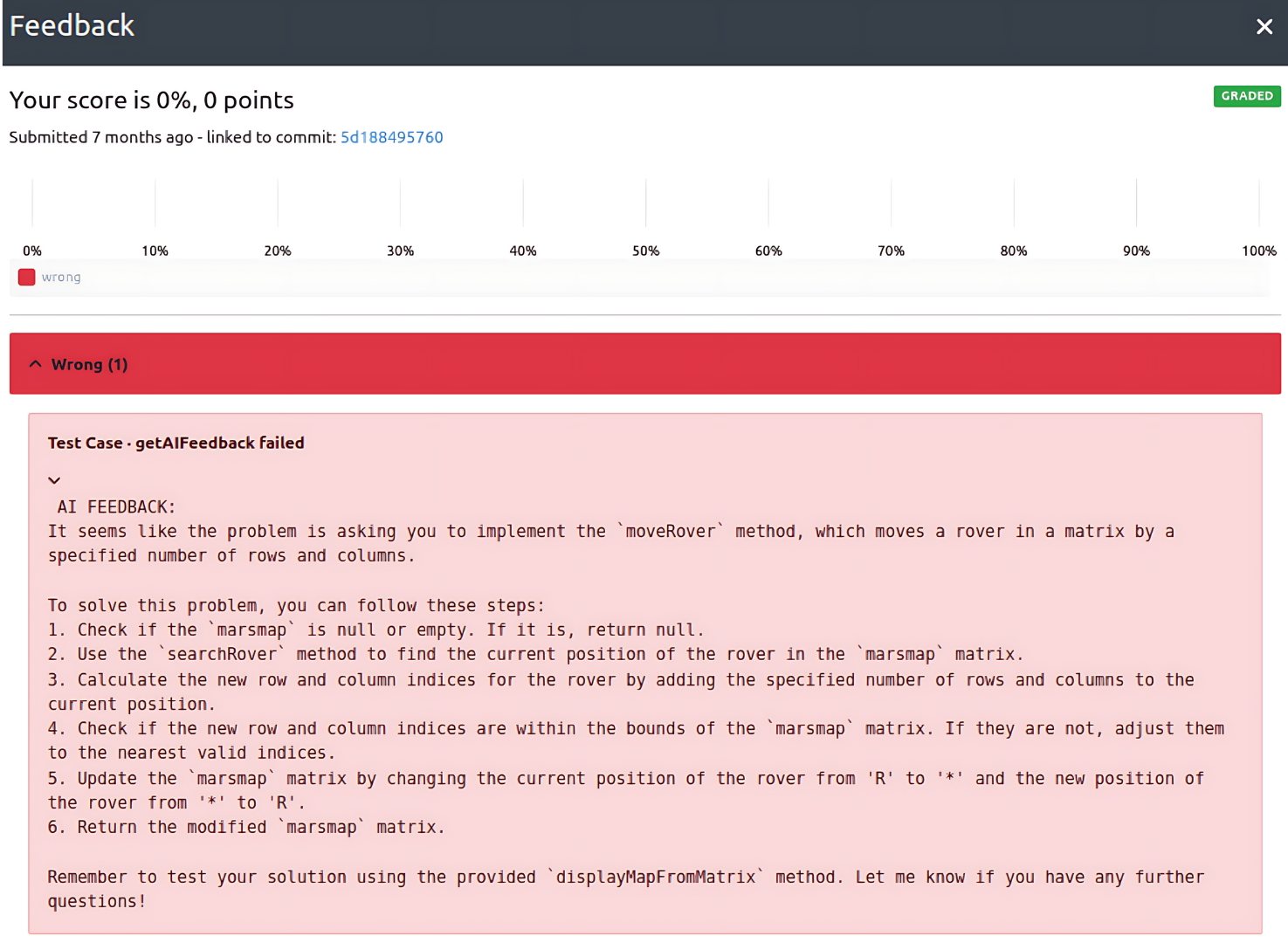}
    \caption{Example of AI-Generated Feedback}
    \label{fig:aifeedback}
\end{figure}

We chose not to enable custom interactions with the AI model for several reasons:

\begin{enumerate}[itemsep=0pt, topsep=0pt]
\item \textbf{Ease of Use:} Predefined prompts streamlined the interaction process, removing the need for students to formulate questions and ensuring \revmod{more} consistency in feedback.
\item \textbf{Controlled Environment:} This also prevented students from trying to trick the system by engineering prompts that tell the system to reveal solutions.
\item \textbf{Quality Control:} The static prompt allowed us to pre-evaluate responses for relevance and quality. 
\item \textbf{Data Privacy:} Restricting custom interactions minimized the risk of students inadvertently sharing personal information in prompts.
\end{enumerate}

\subsection{Data Cleaning}
\label{subsec:data_cleaning}

We applied a strict data cleaning process across all collection points. Initially, any student who did not complete a data collection point was removed from the dataset to maintain consistency. For instance, if a student did not submit a solution for Exercise 2, all data associated with their other exercises and questionnaires were removed. 

For outlier detection and removal, we applied the interquartile~range~(IQR)~method, commonly used when the distribution of data deviates from normality \cite{tukey1977exploratory}. Specifically, we calculated the first quartile (Q1) as well as the third quartile (Q3) for each key variable and determined the IQR as the difference between Q3 and Q1. Any data points falling below $\text{Q1} - 1.5\cdot\text{IQR}$ or above $\text{Q3} + 1.5\cdot\text{IQR}$ were flagged as outliers and subsequently removed. This approach helped reduce skew and ensure that the cleaned dataset accurately reflected typical student performance and feedback responses.

\subsection{Data Analysis}
\label{subsec:data_analysis}

To investigate the impact of each feedback type on student performance and experiences, we conducted a series of statistical analyses.

\begin{itemize}[itemsep=0pt, topsep=0pt]
    \item \textbf{Descriptive Statistics:} For each feedback type, we calculated the mean, median, and standard deviation, providing a foundational understanding of central tendencies and variability within each group.
    \item \textbf{Normality Test:} To assess data distribution, we used the Shapiro--Wilk~test~\cite{shapiro1965analysis}, which evaluates whether the data follows a normal distribution. The results of this test indicated \( p \)-values well below the significance threshold (\( p < 0.01 \)), confirming that the data deviates significantly from normality and validating our selection of non-parametric tests for subsequent analyses.
    \item \textbf{Kruskal-Wallis H-test:} Given the lack of normality and heterogeneous variances, we applied the Kruskal--Wallis~test~\cite{kruskal1952use}, a robust non-parametric test, to examine significant differences in correctness scores across feedback types. This test is particularly suited for comparing independent samples in non-normally distributed data.
    \item \textbf{Post-hoc Dunn's Test:} When the Kruskal--Wallis~test indicated statistically significant differences, we conducted Dunn's~test \cite{dunn1964multiple} with Bonferroni~correction to perform pairwise comparisons and identify specific group differences. This post-hoc test determined which feedback types were statistically different from each other in terms of student performance.
\end{itemize}

To enhance interpretation, we visualized the data using box plots:

\begin{itemize}
    \item \textbf{Box Plots:} Box plots were used to provide a clear visual summary of each feedback type's data distribution, including the median, quartiles, and outliers. This visualization facilitated comparison across feedback types, highlighting variations in performance and feedback effectiveness.
\end{itemize}

Overall, the combination of non-parametric statistical tests and visual analysis ensured a comprehensive examination of the data, enabling reliable insights into the effects of feedback types on student outcomes.

\begin{figure}[b!]
    \centering
    \includegraphics[width=0.80\linewidth]{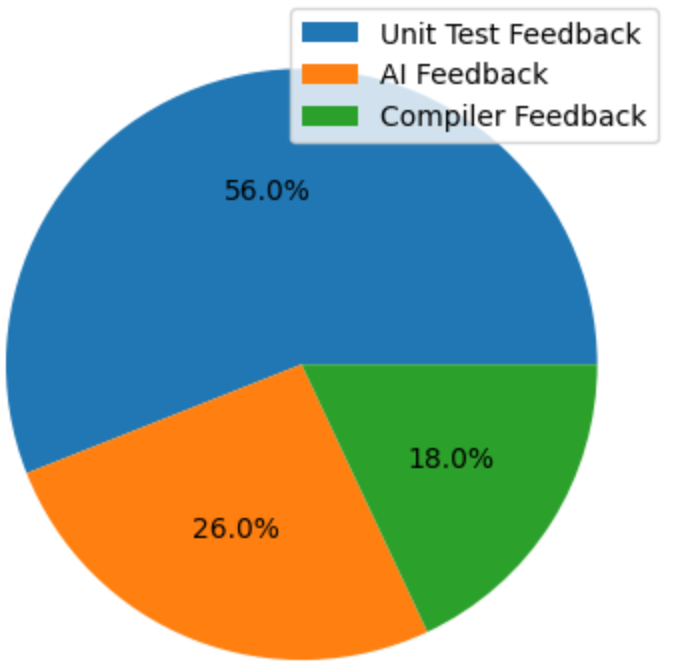}
    \caption{Best Feedback Type According to Students}
    \label{fig:best_feedback}
\end{figure}

\revmod{For the open-ended survey answers (DC4), we applied a simple thematic analysis. All free-text statements were collected and read in full. If a statement occurred only once, we still report. If multiple students made similar statements, we grouped them into themes and counted the number of times they appeared. Figure \ref{fig:best_feedback} in Section~\ref{sec:perceived_utility} therefore shows only descriptive statistics of preferences, while the narrative reports themes and frequencies derived from this analysis.}


\section{Participants}
\label{sec:demographics}
A total of $212$ university students participated in the survey ($74$ participants from the first university, and $138$ participants from the second university). Among them, $170$ identified as male and $42$ as female, with ages ranging from $17$ to $62$ years ($\text{mean} = 22.15\text{ years}$, $\text{SD} = 4.11\text{ years}$, $\text{median} = 21\text{ years}$). Four participants did not specify their age. Regarding educational qualifications, the majority ($186$ participants) reported holding an A-level degree as their highest qualification. Other qualifications included a bachelor's degree ($15$ participants), a master's degree (5 participants), and a completed apprenticeship ($2$ participants). Additionally, four participants indicated holding other unspecified qualifications. 

\begin{figure}[H]
    \centering
\includegraphics[width=0.99\linewidth]{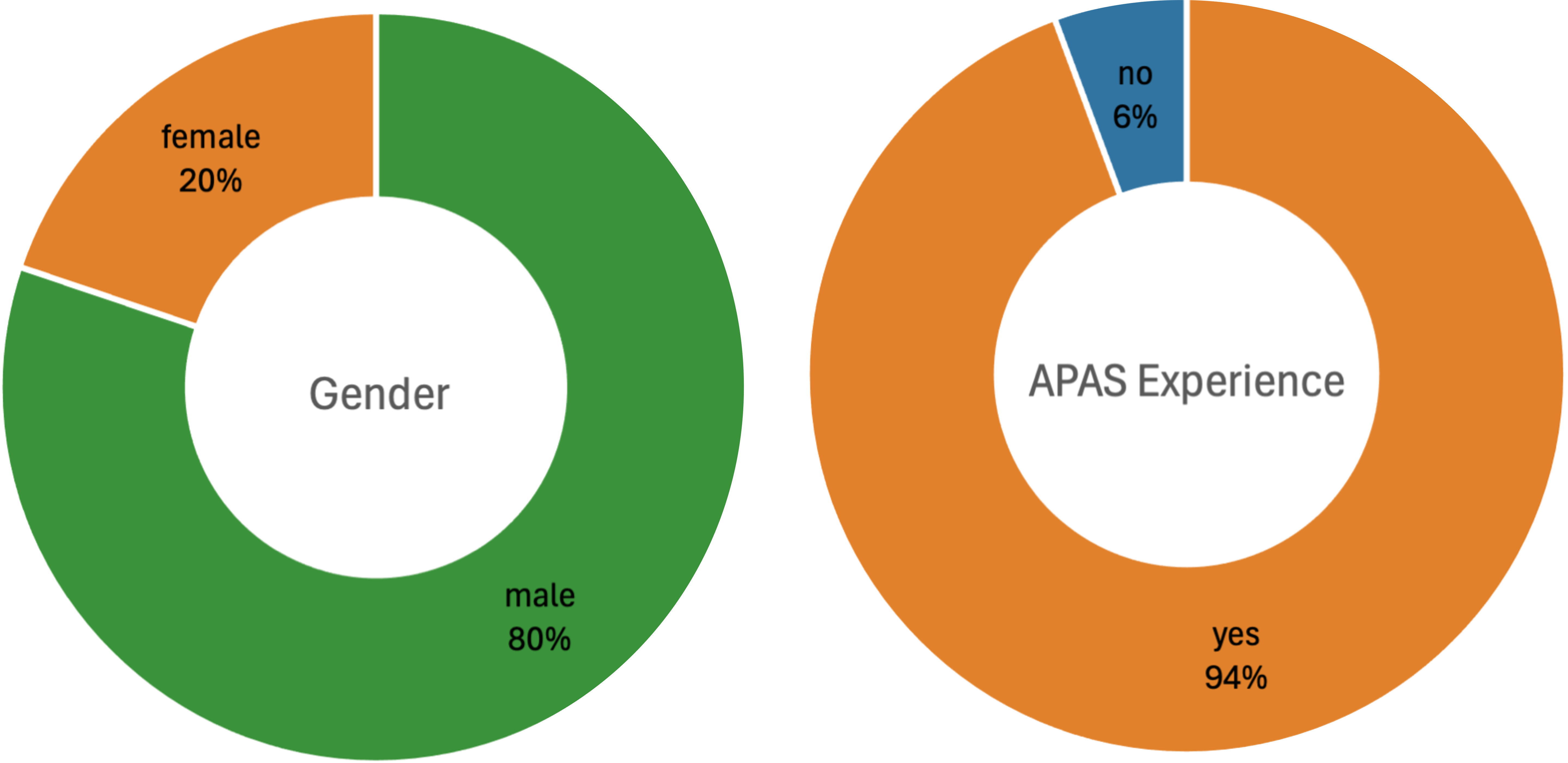}
    \caption{Overview of Students' Gender Distribution and Experience with APASs}
    \label{fig:gender}
\end{figure}

More than half of the participants ($118$ students) were enrolled in bachelor's programs in computer science, while $54$ students were enrolled in the bachelor's program in business informatics. One participant reported being in a master's program in business informatics. Other fields of study included business economics (three students) and statistics and data science ($14$ students). $22$ participants were enrolled in other programs, $12$ of whom stated they were pursuing an extension program in computing alongside their main program. Three were simultaneously pursuing a bachelor's degree in computer science and another bachelor's degree, while one participant was training as a student teacher in computing. Two participants were studying economics, and the remaining four were enrolled in medicine, mechatronics, psychology, and an extraordinary study program, respectively. 

At the time of the study, the majority of participants were in their first or second semester, with $69$ students in their first semester and $98$ in their second. The remaining participants were distributed across various semesters: seven in the third semester, $14$ in the fourth semester, three in the fifth semester, and $12$ in the sixth semester. Additionally, nine participants were in higher semesters, including three in the seventh semester and six in the eighth semester. Students were also asked how many times they had attended the introductory course in programming already. Of the $212$ participants, $189$ ($89\%$) were attending the course for the first time, $20$ ($10\%$) for the second time, and $3$ ($1\%$) for the third time. Programming experience among participants ranged from none ($0$ years) to $11$ years ($\text{mean} = 2.10\text{ years}$, $\text{SD} = 2.30\text{ years}$, $\text{median} = 1\text{ year}$), with five students not specifying their level of experience. 


\begin{figure}[h!]
  \centering
  
  \begin{subfigure}{0.9\columnwidth}
    \centering
    \includegraphics[width=\linewidth]{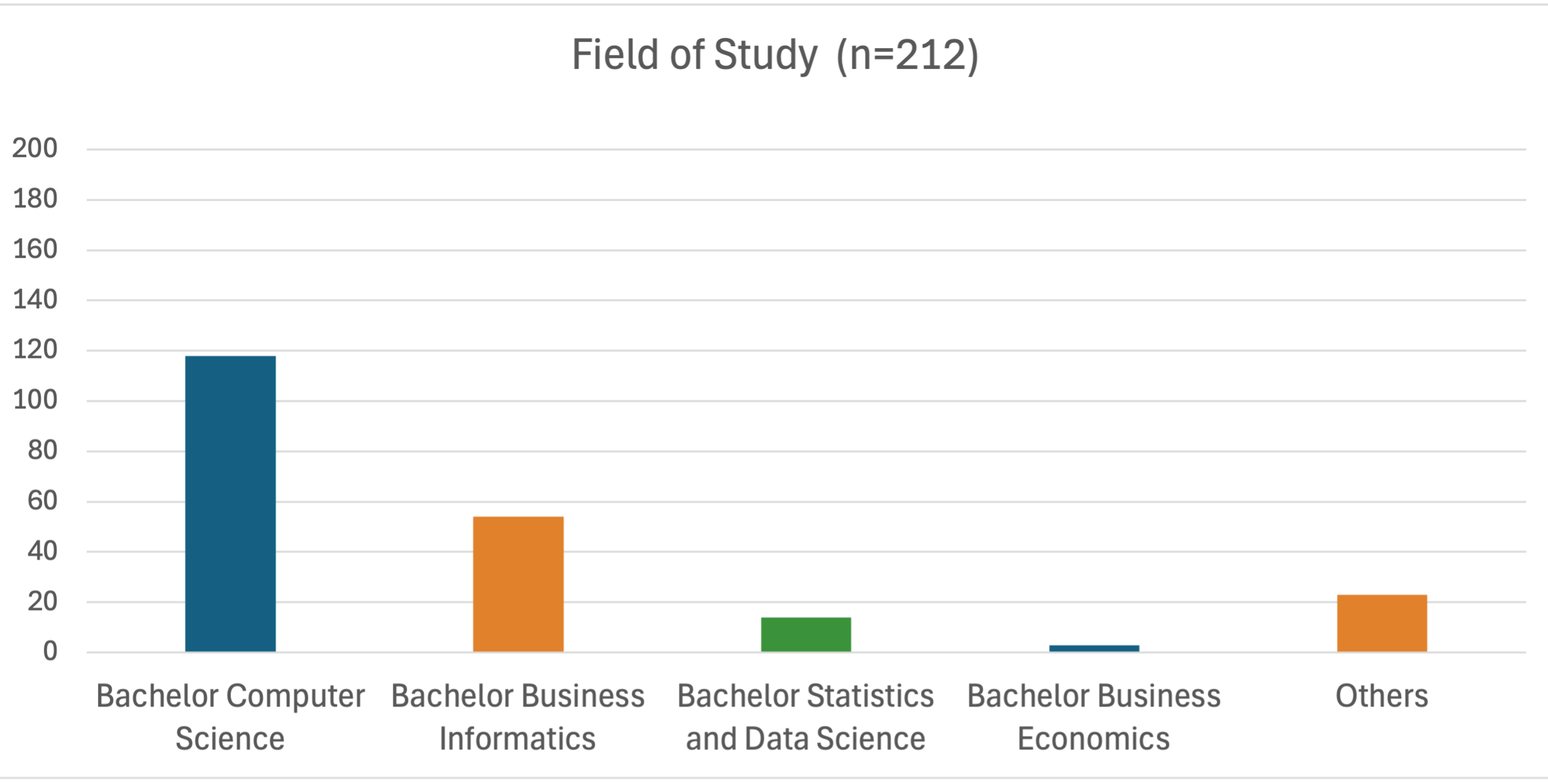}
    \label{fig:sub1}
  \end{subfigure}
  
  \vspace{0.2em} 
  
  \begin{subfigure}{0.9\columnwidth}
    \centering
    \includegraphics[width=\linewidth]{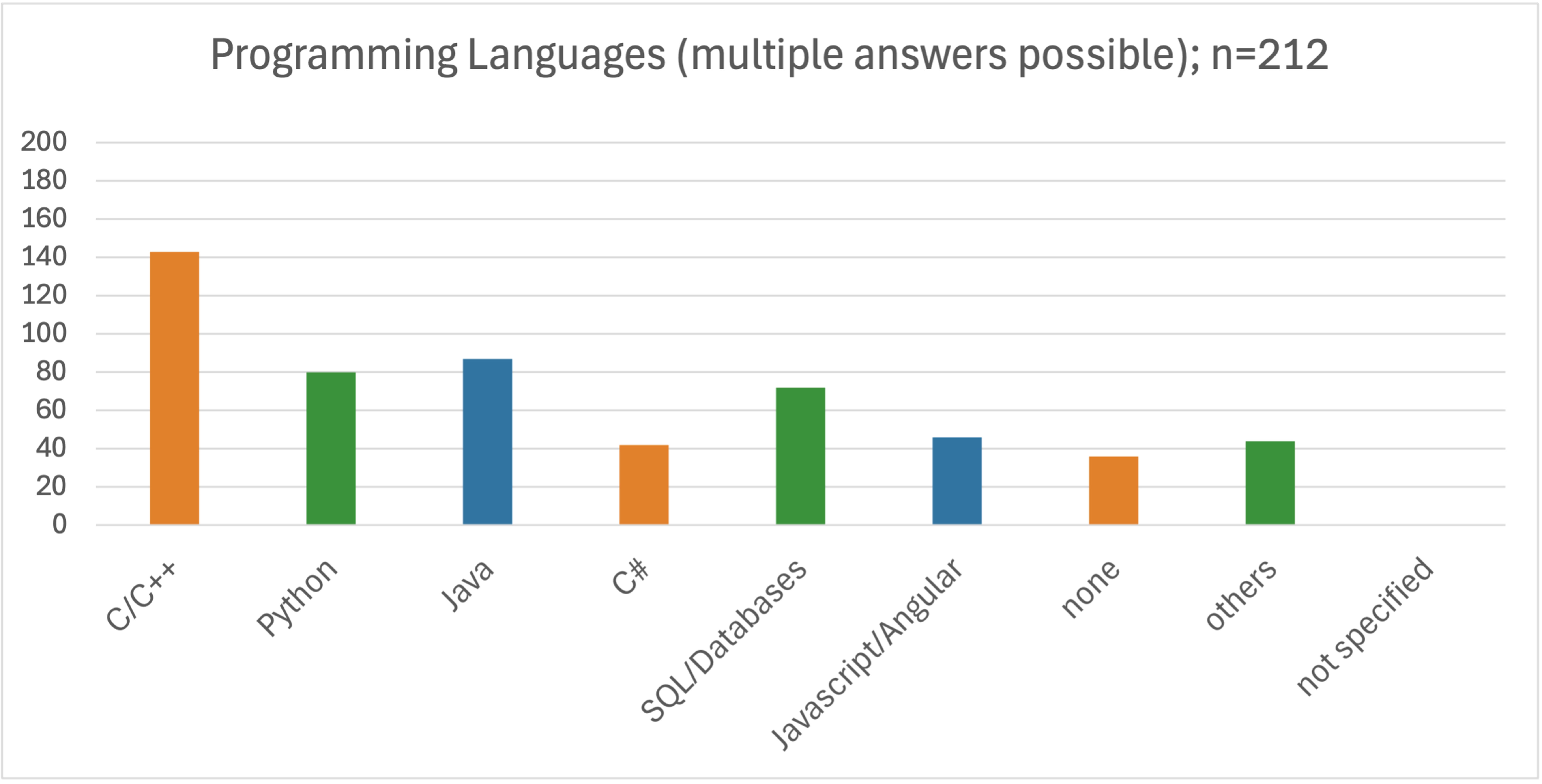}
    \label{fig:sub2}
  \end{subfigure}

 \caption{Students' experienced programming languages and their current study programs.}
  \label{fig:participants}
  \vspace{-0.9em}
\end{figure}

When asked about \revmod{prior} familiarity with programming languages, the most commonly cited language was C/C++ ($143$ respondents), followed by Java ($87$ respondents), Python ($80$ respondents), and SQL or database systems ($72$ respondents). C\# and JavaScript/Angular were also frequently mentioned, with $42$ and $46$ respondents, respectively. Notably, $36$ participants indicated having no experience with programming languages, while $44$ reported proficiency in other languages, collectively naming $25$ additional languages in $77$ responses. 
Regarding familiarity with the APAS used in the study, most participants ($200$ students, $94\%$) reported prior experience with the system. Only $12$ participants ($6\%$) indicated that they had not used the system before the experiment (cf. \citefig{gender}). \citefig{participants} provides an overview of the different fields of study and programming languages in which students have indicated experience.

\section{Part 1: Student Perceptions of Feedback Types}
\label{sec:part1}
This section presents findings pertaining to \textbf{RQ1}, which explores students' perceptions of different feedback types regarding quality and usefulness (RQ1.1), and overall perceived utility (RQ1.2).

\subsection{TAM -- Quality and Usefulness of Feedback Types}
\label{sec:quality_usefulness}

\begin{figure}[H]
    \centering
    \includegraphics[width=0.99\linewidth]{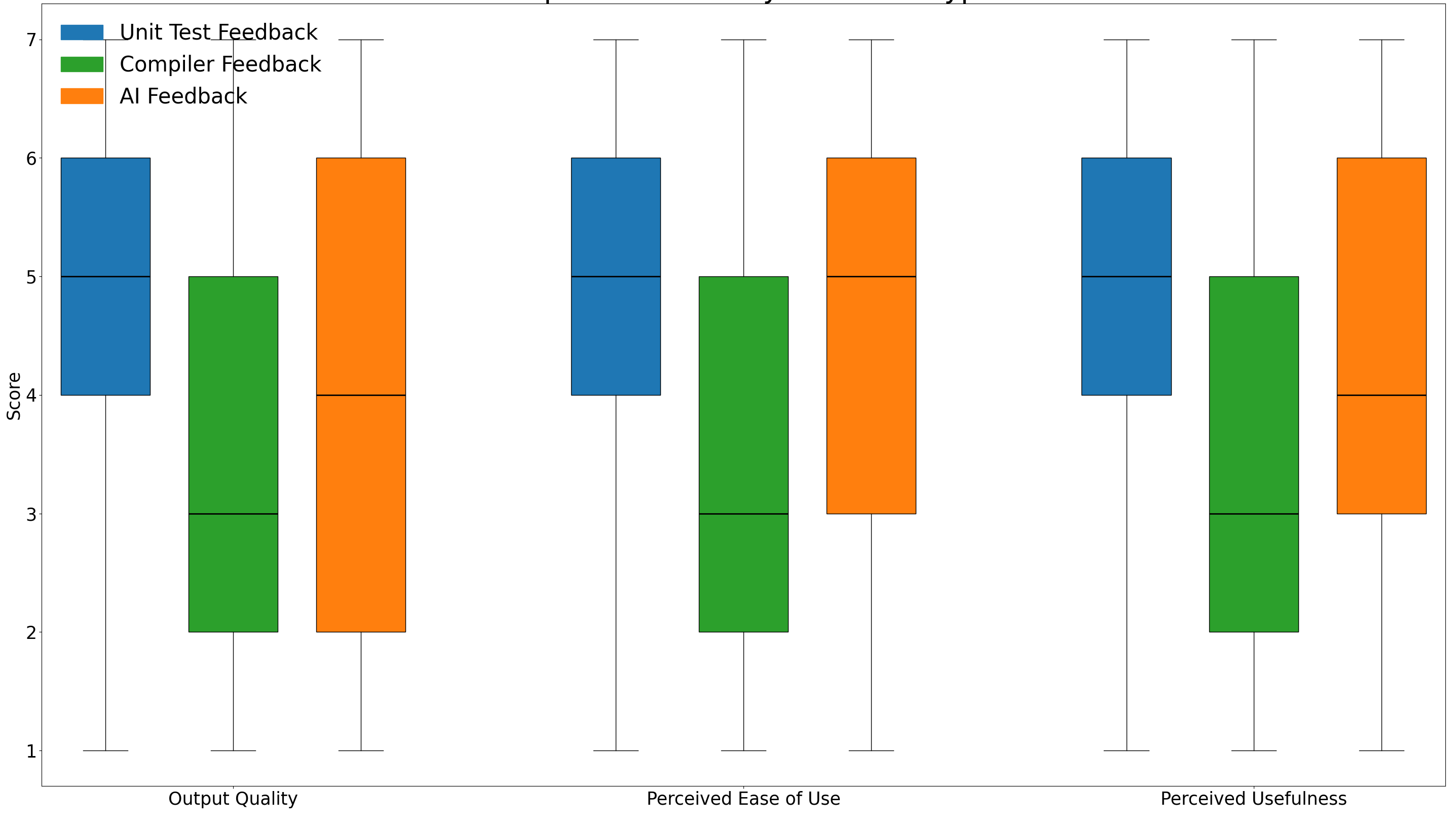}
    \caption{Comparative Scores for Quality and Usefulness by Feedback Type}
\label{fig:feedback_quality_boxplot}
\end{figure}

Fig.~\ref{fig:feedback_quality_boxplot} shows the comparative scores for quality and usefulness of feedback types as rated by students, illustrating the distribution of scores for AI, Compiler, and Unit Test Feedback. This data is based on analyzing results from the three TAM questionnaires (DC3) handed out after each exercise. For each of the three constructs (PU, EU, OQ), we analyzed the results for each of the three feedback types.

\textbf{Statistical Analysis:} The reliability of the scale was assessed using Cronbach's~$\alpha$, Guttman's~$\lambda_{6}$, and average inter-item correlation. Cronbach's~$\alpha$ was approximately $0.93$, and Guttman's~$\lambda_{6}$ was $0.97$, indicating excellent internal consistency. The average inter-item correlation was $0.34$, suggesting good internal consistency without excessive redundancy. The $95\%$ confidence intervals for the reliability coefficient were narrow and above $0.90$, further reinforcing the reliability of the scale.
For all items, removing any one does not change the overall alpha significantly. Most of the values remain at approximately $0.92$, which indicates that each item is contributing positively to the scale's reliability.

\textbf{TAM Results:} 
For AI Feedback, the mean ($\mu$) is $4.2$, the median (M) is $4.5$, and the standard deviation ($\sigma$) is $1.9$.
For Compiler Feedback, the mean ($\mu$) is $3.4$, the median (M) is $3.3$, and the standard deviation ($\sigma$) is $1.8$.
For Unit Test Feedback, the mean ($\mu$) is $4.8$, the median (M) is $5.0$, and the standard deviation ($\sigma$) is $1.7$.

The comparison of means indicates that Unit Test Feedback was rated the highest, followed by AI Feedback, with Compiler Feedback receiving the lowest ratings. This suggests that students found Unit Test Feedback most useful and consistent, while AI Feedback had more variability in responses.

For all three examined TAM constructs, i.e., Perceived Usefulness, Perceived Ease of Use, and Output Quality, across the three feedback types AI, Compiler, and Unit Test, the Shapiro-Wilk test $p$-values are well below the significance level of $0.001$, indicating that the data for each construct does not follow a normal distribution. As a result, the Kruskal-Wallis test, a non-parametric alternative to ANOVA, was used for further comparison between feedback types.

The Kruskal-Wallis test ($\chi^2 = 535.29$, $p < .001$) showed statistically significant differences between the three feedback types. Pairwise comparisons using Dunn's test with Bonferroni correction revealed the following results seen in Table \ref{tab:stat_analysis_quality}:

\begin{table}[H]
    \centering
    \begin{tabular}{|l|c|c|}
        \hline
        \rowcolor{customgray} \textbf{Comparison} & \textbf{p-adj} & \textbf{Significant} \\ \hline
        \cellcolor{customgray} AI Feedb. vs. Compiler Feedb. & $< .001$ & Yes \\ \hline
        \cellcolor{customgray} AI Feedb. vs. Unit Test Feedb. & $< .001$ & Yes \\ \hline
        \cellcolor{customgray} Compiler Feedb. vs. Unit Test Feedb. & $< .001$ & Yes \\ \hline
    \end{tabular}
    \vspace{10pt}
    \caption{Post-hoc Dunn's test results (p-adj) for the quality and usefulness scores by feedback type.}
    \label{tab:stat_analysis_quality}
\end{table}

The results indicate significant differences between all pairs of feedback types. AI Feedback was rated significantly higher than Compiler Feedback but lower than Unit Test Feedback. Compiler Feedback received the lowest ratings.

\subsection{Perceived Utility of Feedback Types}
\label{sec:perceived_utility}
In addition to results obtained from the TAM questionnaires, we collected additional feedback using the post-survey questionnaire (DC4), to gain additional insights into, for example, why students deemed a feedback type particularly helpful and useful, or why not.
A total of 85 of the 212 students who initially participated in the experiment also participated in the post-survey questionnaire. The survey asked students about their preferred feedback form. As illustrated in~\citefig{best_feedback}, 56\% of participants favored the Unit Test Feedback, followed by 26\% that favored AI Feedback, while Compiler Feedback, 18\%, was the least preferred.

We further asked the students to provide feedback regarding the different feedback forms in the experiment. In the following, we summarize the students' responses.
Ten students across the universities found the AI Feedback not helpful, with issues such as providing incorrect or hallucinated error messages. AI Feedback was initially helpful for most students, but became less useful when it provided confusing or incorrect suggestions once the code was mostly correct. 
Two students found AI Feedback beneficial for optimization suggestions and learning new implementation techniques. Students also liked its ability to adapt to the code and provide personalized feedback.
One student also noted that AI Feedback can sometimes provide quicker solutions, but it may reduce critical thinking by offering step-by-step solutions to students.
Suggestions for improving the AI Feedback in the future included improving the clarity of the feedback, providing more concrete responses, and the ability to ask questions directly.

Six students found the Compiler Feedback helpful for identifying and correcting trivial errors, while four found it confusing and not helpful. In general, Compiler Feedback was valued for identifying and correcting simple errors. It was less helpful for more complex issues due to unclear messages.
Compiler Feedback often caused more confusion than assistance, with five students having problems understanding and fixing errors within the given timeframe.
For three students, Compiler Feedback was straightforward and aligned well with basic debugging practices.

Unit tests were generally seen as the most effective form of feedback for achieving correct results. Students liked the correctness verification provided by unit tests, though some found the feedback impractical under time constraints.
Students found unit tests easier to understand and more helpful in identifying specific issues.
They reported that unit tests provided clear feedback on what was expected versus what was produced, also highlighting exactly where problems occurred and supporting error detection and correction.
Many students were familiar with unit tests from previous experiences, and they found them efficient for debugging.
The students noted that unit tests allow for flexibility in finding solutions without imposing specific methods, fostering creativity and independent problem-solving.

In the survey, we also asked the students to report on other forms of feedback they would find helpful.
Students requested that the compiler should further provide immediate feedback directly to save time, rather than having to wait for a complete evaluation from the Artemis environment.

It was also mentioned that the Artemis programming environment should use visual aids such as red underlines to highlight errors in the code, similar to other editors like Visual Studio Code.
A combination of feedback forms was seen as the most helpful by several students. They suggested a combination of Unit Tests and AI Feedback. Also, AI Feedback should stop providing suggestions once the code is correct to prevent unnecessary changes. A combination of Unit Tests and Compiler Feedback was regarded as helpful as well, to see both what is expected for submission and to enable print statement-based debugging for personalized outputs.

In the survey, we also asked the students to give general feedback regarding the experiment setup.
In general, the students noted that they did not encounter major problems during the experiment. The Artemis programming environment was criticized for being slow by some students. A recurring topic across the universities was the need for more time to engage more deeply with the different feedback mechanisms. 

\begin{findingsBox}
\textbf{Main Findings for RQ1}:
\begin{itemize}
    \item Unit Test Feedback received the highest ratings for quality and usefulness.
    \item AI Feedback was rated higher than Compiler Feedback but lower than Unit Test Feedback.
    \item Compiler Feedback, while helpful for trivial errors, was often unclear for more complex issues.
    \item Students preferred Unit Test Feedback for clarity and effectiveness, and some suggested combining it with AI Feedback for optimal results.
    \item AI Feedback provided useful optimization tips and personalized guidance, but occasionally gave incorrect or confusing suggestions.
\end{itemize}
\end{findingsBox}

\section{Part 2: Impact of Feedback Types on Student Performance}
\label{sec:part2}
In this section, we present results from our quantitative analysis of the collected exercise data (DC2). Addressing  \textbf{RQ2}, we focus on how different feedback forms influence students' performance (RQ2.1) and problem-solving strategies (RQ2.2).

\subsection{Feedback Types and Student Performance}
\label{sec:performance}

\citefig{comparative_final_correctness} provides an overview of the comparative final correctness scores per student by feedback type after running the extra set of unit tests on the students' solutions. The box plots illustrate the distribution of final correctness scores across different feedback types, providing insights into how each form of feedback influences the correctness of students' final submissions.

\textbf{Statistical Analysis:} The Shapiro-Wilk test for normality yielded a statistic of 0.824 with a p-value < \(1 \times 10^{-25}\), indicating that the data significantly deviates from normality. Therefore, non-parametric tests were used.

\textbf{Score Results:} For AI Feedback, the mean ($\mu$) is 65.44, the median (M) is 85.70, and the standard deviation ($\sigma$) is 38.67. For Compiler Feedback, the mean ($\mu$) is 41.41, the median (M) is 23.80, and the standard deviation ($\sigma$) is 40.45. For Unit Test Feedback, the mean ($\mu$) is 57.52, the median (M) is 70.00, and the standard deviation ($\sigma$) is 37.77.

The comparison of means shows that AI Feedback has the highest mean score, followed by unit test support, and no support exhibits the lowest mean score. This suggests that students with AI Feedback tend to achieve higher correctness in their final submissions. 
To verify whether there is a significant difference in the final correctness scores depending on the feedback type, we performed the Kruskal-Wallis H-test and Dunn's post-hoc test.

\begin{table}[H]
    \centering
    \begin{tabular}{|l|c|c|c|}
        \hline
        \rowcolor{customgray} \textbf{Comparison} & \textbf{p-adj} & \textbf{Significant} \\ \hline
        \cellcolor{customgray} AI Feedb. vs. Compiler Feedb. & $< .001$ & Yes \\ \hline
        \cellcolor{customgray} AI Feedb. vs. Unit Test Feedb. & 0.035 & Yes \\ \hline
        \cellcolor{customgray} Compiler Feedb. vs. Unit Test Feedb. & $< .001$ & Yes \\ \hline
    \end{tabular}
    \vspace{10pt}
    \caption{Post-hoc Dunn's test results for the final correctness scores by feedback type}
    \label{tab:stat_analysis_correctness}
\end{table}

\begin{figure}[H]
    \centering
    \includegraphics[width=0.99\linewidth]{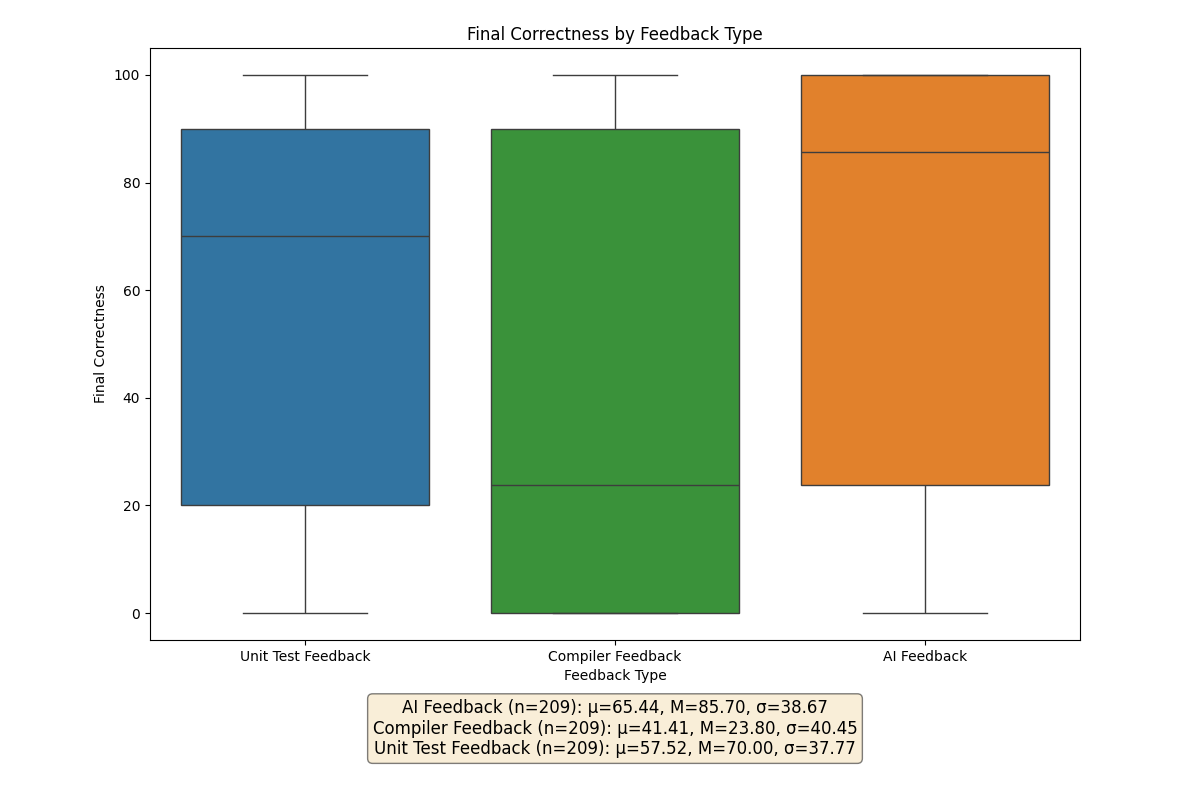}
    \caption{Comparative Final Correctness Scores by Feedback Type}
    \label{fig:comparative_final_correctness}
\end{figure}

\subsection{Problem-Solving Approaches with Different Feedback Types}
\label{sec:problem_solving}
This subsection analyzes the effect of different feedback types on students’ problem-solving strategies and methods, examining the number of submissions and the time lapsed between submissions.

\citefig{comp_attemps} shows the comparative number of submissions per student, again grouped by the respective feedback type. Illustrating the distribution of the number of attempts made by students across different feedback types provides interesting insights into how each form of feedback influences the persistence and iteration behavior of students during programming tasks.

After applying IQR filtering, the data size was reduced slightly (from 640 to 624), ensuring the exclusion of extreme outliers. The Shapiro-Wilk test for normality yielded a statistic of 0.97 with a p-value < \(1 \times 10^{-10}\), indicating that the data significantly deviates from normality. Therefore, non-parametric tests were used.

For AI Feedback Support, the mean ($\mu$) is 8.95, the median (M) is 8.00, and the standard deviation ($\sigma$) is 3.41. 

For Compiler Feedback Support, the mean ($\mu$) is 7.98, the median (M) is 8.00, and the standard deviation ($\sigma$) is 3.37. 

For Unit Test Feedback Support, the mean ($\mu$) is 8.08, the median (M) is 8.00, and the standard deviation ($\sigma$) is 3.37. 

The comparison of means shows that the means of the three groups are close, with AI Feedback having the highest mean attempts, followed by unit test support, and no support having the lowest mean. This suggests that students with AI Feedback tend to iterate more on their solutions. The spread and outliers indicate that while there are individual differences in persistence, the overall impact of feedback type on the number of attempts is consistent. The medians suggest that the central tendency of attempts is similar across all support types, with students typically making around 8 attempts regardless of the support type.

The Kruskal-Wallis H-test (\(H=9.299\), \(p=0.009\)) confirmed a statistically significant difference in the distributions of attempts across feedback types. Dunn's post-hoc test with Bonferroni correction was applied for pairwise comparisons (Table \ref{tab:stat_analysis_updated}).

\begin{table}[H]
    \centering
    \begin{tabular}{|l|c|c|}
        \hline
        \rowcolor{customgray} \textbf{Comparison} & \textbf{p-adj} & \textbf{Significant} \\ \hline
        \cellcolor{customgray} AI Feedb. vs. Compiler Feedb. & 0.016 & Yes \\ \hline
        \cellcolor{customgray} AI Feedb. vs. Unit Test Feedb. & 0.042 & Yes \\ \hline
        \cellcolor{customgray} Compiler Feedb. vs. Unit Test Feedb. & 1.0000 & No \\ \hline
    \end{tabular}
    \vspace{10pt}
    \caption{Post-hoc Dunn's test results for the number of attempts by feedback type}
    \label{tab:stat_analysis_updated}
\end{table}

The results indicate significant differences in attempts between AI Feedback and both Compiler Feedback and Unit Test Feedback. The effect size (Eta Squared = 1.55) suggests that feedback type has a meaningful impact on students' iteration behavior. This suggests that students receiving AI Feedback use more attempts to solve an exercise compared to those receiving no support or unit test support. This finding is important as it implies that AI Feedback might encourage more iteration and exploration, potentially leading to a deeper understanding of the programming tasks.

\begin{figure}[b!]
    \centering
    \includegraphics[width=0.99\linewidth]{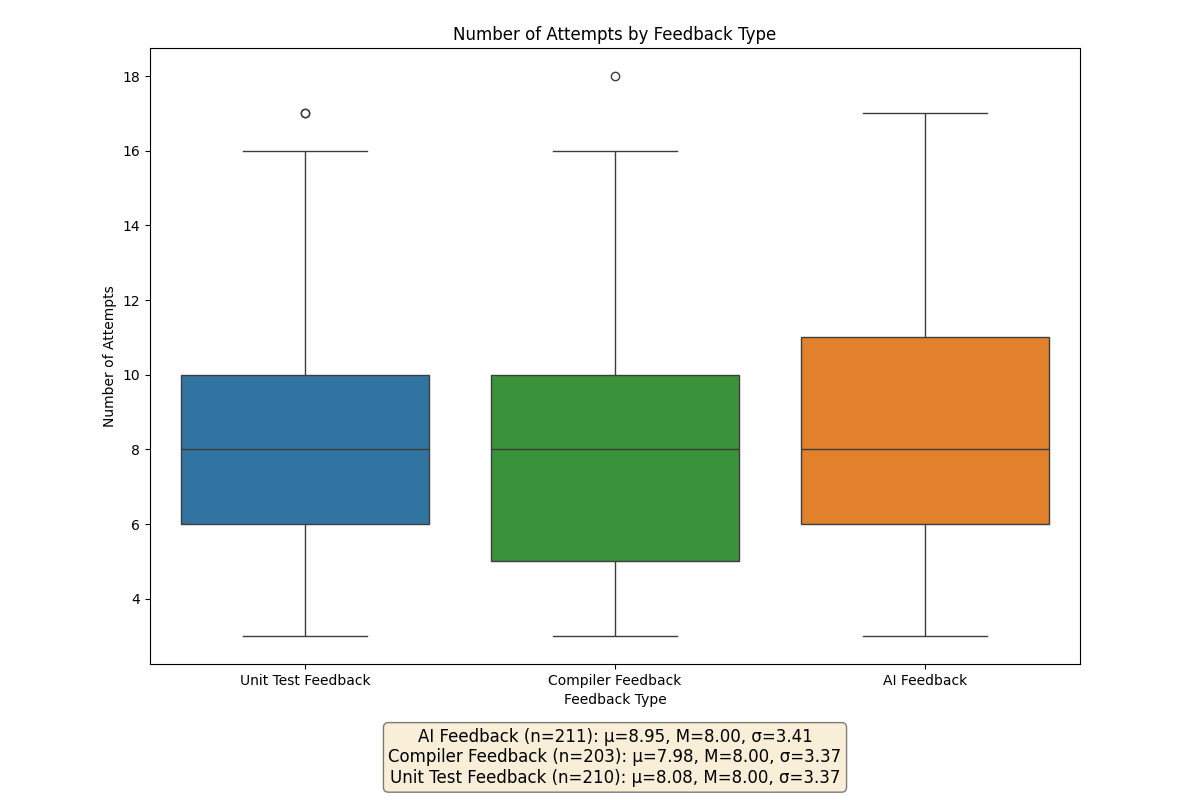}
    \caption{Comparative Number of Attempts per Student by feedback type}
    \label{fig:comp_attemps}
\end{figure}

\begin{figure}[th!]
    \centering
    \includegraphics[width=0.99\linewidth]{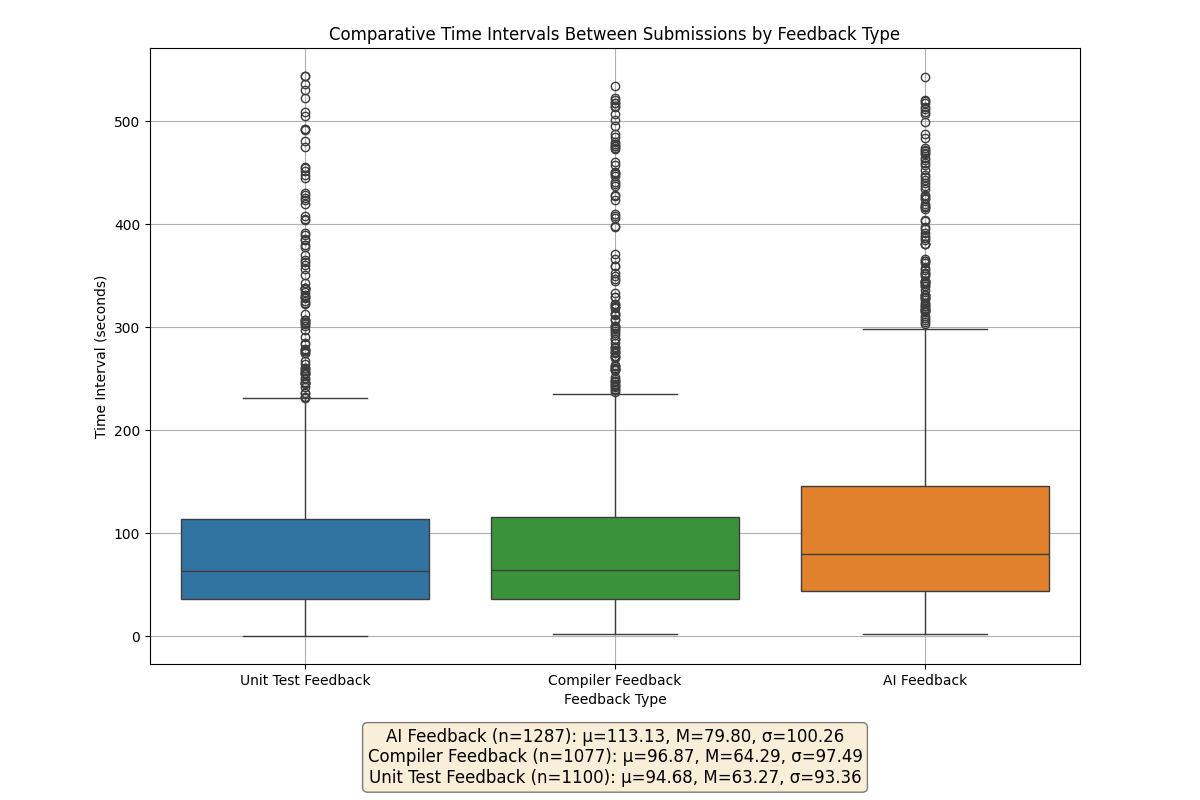}
    \caption{Comparative Time Intervals Between Submissions by feedback type}
{}
    \label{fig:comp_time_intervals}
\end{figure}

In the following, we analyzed the effect of different feedback types on the time between submissions by students.
After applying IQR filtering, the dataset size was reduced from 4199 to 3464, ensuring the exclusion of extreme outliers. The Shapiro-Wilk test for normality yielded a statistic of 0.766 with a p-value of 0.0, indicating that the time interval data significantly deviates from normality. Thus, non-parametric tests were applied.
Fig. \ref{fig:comp_time_intervals} presents the comparative time intervals between submissions by feedback type. 

For AI Feedback, the mean ($\mu$) is 113.13 seconds, the median (M) is 79.80 seconds, and the standard deviation ($\sigma$) is 100.26 seconds. 

For Compiler Feedback, the mean ($\mu$) is 96.87 seconds, the median (M) is 64.29 seconds, and the standard deviation ($\sigma$) is 97.49 seconds. 

For Unit Test Feedback, the mean ($\mu$) is 94.68 seconds, the median (M) is 63.27 seconds, and the standard deviation ($\sigma$) is 93.36 seconds. 

The comparison of means indicates that AI Feedback leads to the longest average time intervals between submissions, followed by no support, with unit test support resulting in the shortest intervals.
To determine whether these differences are statistically significant, a Kruskal-Wallis H-test (\(H=49.17\), \(p=2.10 \times 10^{-11}\)) was performed. Post-hoc pairwise comparisons using Dunn's test were conducted, and the results are presented in Table \ref{tab:stat_analysis_time_intervals}.

\begin{table}[H]
    \centering
    \begin{tabular}{|l|c|c|}
        \hline
        \rowcolor{customgray} \textbf{Comparison} & \textbf{p-adj} & \textbf{Significant} \\ \hline
        \cellcolor{customgray} AI Feedb. vs. Compiler Feedb. & $< .001$ & Yes \\ \hline
        \cellcolor{customgray} AI Feedb. vs. Unit Test Feedb. & $< .001$ & Yes \\ \hline
        \cellcolor{customgray} Compiler Feedb. vs. Unit Test Feedb. & 1.0000 & No \\ \hline
    \end{tabular}
    \vspace{10pt}
    \caption{Post-hoc Dunn's test results for time intervals between submissions by feedback type.}
    \label{tab:stat_analysis_time_intervals}
\end{table}

The results indicate significant differences between AI Feedback and both Compiler Feedback and Unit Test Feedback. However, the difference between Compiler Feedback and Unit Test Feedback is not statistically significant. The effect size (Eta Squared = 8.19) suggests a substantial impact of feedback type on the time intervals.

\begin{figure}[th!]
    \centering
    \includegraphics[width=0.99\linewidth]{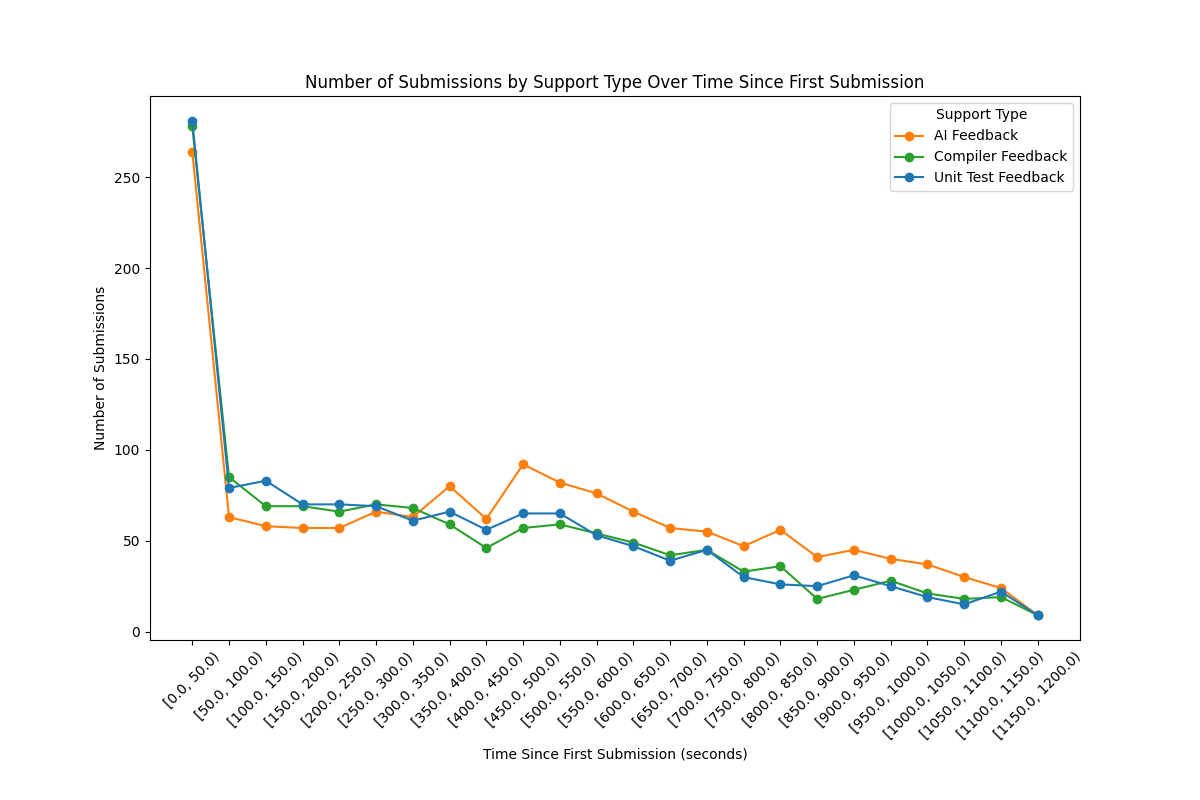}
    \caption{Number of Submissions by feedback type Over Time Since First Submission}
    \label{fig:submissions_over_time}
\end{figure}

The analysis reveals that AI Feedback leads to longer and more variable intervals between submissions, suggesting that students take more time to iterate and refine their work. Compiler feedback results in shorter intervals, indicating quicker successive submissions. Unit Test Feedback falls in between, with variability comparable to AI Feedback.
Fig. \ref{fig:submissions_over_time} shows the number of submissions by feedback type over time since the first submission. The line plots illustrate how the frequency of submissions decreases over time for each feedback type, providing insights into the submission patterns and persistence of students under different forms of feedback.
\revmod{Submissions with Compiler Feedback are initially more frequent but exhibit a steeper decline over time compared to the other feedback types. In contrast, AI Feedback shows a lower initial submission rate but maintains a relatively higher level across time, suggesting sustained usefulness.}

\begin{findingsBox}
\textbf{Main Findings for RQ2}:
\begin{itemize}
    \item Students receiving AI Feedback achieved the highest correctness scores, significantly outperforming those with Unit Test Feedback or Compiler Feedback support.
    \item Unit Test Feedback also improved correctness scores compared to no support, though not as strongly as AI Feedback.
    \item \revmod{AI Feedback led to students making more attempts and taking longer intervals between submissions}.
    \item Compiler Feedback led to quicker successive submissions but lower correctness and less sustained problem-solving effort over time.
\end{itemize}
\end{findingsBox}

\section{Discussion}
\label{sec:discussion}

The results of this study show that not all feedback forms are equally helpful to students. Traditional compiler feedback, while quick to point out simple syntax mistakes, often leaves students struggling when it comes to more complex issues. In contrast, both unit tests and AI-generated feedback offered clearer guidance and supported students in producing more accurate solutions.\\

\noindent\textbf{Familiarity with Unit Tests:} 
\noindent An interesting outcome is the discrepancy between students' perceptions and their performances. Students expressed a strong preference for unit test feedback, describing it as clear, straightforward, and useful. However, those who received AI-generated feedback ultimately achieved the highest correctness scores. This suggests that while unit tests may feel more familiar and reassuring, \revmod{AI Feedback encourages deeper engagement because students guided by AI hints took more time between submissions, paused longer before resubmitting, and revisited their code more frequently.}\\ 

\noindent\textbf{More Advanced AI Feedback Mechanisms:}  
\noindent While unit tests provide stable and reliable verification of correctness, they may not inherently foster the depth of reasoning and concept mastery that AI Feedback can inspire. At the same time, although the adaptive and tutor-like guidance of AI Feedback can lead to higher correctness and more sustained problem-solving, it occasionally produces confusing or erroneous suggestions. These inconsistencies underline the importance of improving the reliability of AI-based feedback mechanisms.  

\revmod{As reflected in students' free-text reports in Section~\ref{sec:perceived_utility}, occasional inaccuracies and hallucinations can erode trust even when overall performance improves. Since our setup relied on a fixed, non-interactive prompt yet still exhibited such cases, we view reliability and trust as central concerns for AI feedback in APASs. Building on recent work on adaptive and personalized feedback generation~\cite{bassner2024iris}, more advanced mechanisms are needed to ensure consistency. Potential strategies include grounding responses in verifiable artifacts (e.g., failing unit tests or specific code locations), limiting output length and code generation to reduce error-prone verbosity, and enabling abstention when reliable hints cannot be generated (i.e., the system refrains from giving potentially misleading advice). Additionally, a second-order validation step, where an auxiliary LLM evaluates whether a response is appropriate and consistent with the task, could further filter out misleading outputs and strengthen student trust.}\\  
Beyond these technical safeguards, students themselves emphasized the value of interaction design. Several suggested that a more interactive, chatbot-like interface would allow them to ask follow-up questions and clarify uncertainties in real time, thereby further enhancing both the utility and the trustworthiness of AI feedback.\\

\noindent\textbf{Combined Feedback Mechanisms:} 
\noindent These findings highlight the value of a balanced feedback environment. Unit tests excel at verifying that students meet specific requirements and provide a sense of closure and direction. AI-generated hints, on the other hand, can offer context-sensitive, exploratory support that encourages students to reflect, adapt, and explore alternative solution paths. Integrating both approaches could leverage the clarity and verification of unit tests and the adaptive, personalized scaffolding of AI Feedback, resulting in richer opportunities for skill development and deeper conceptual understanding.

From a pedagogical perspective, integrating AI hints with unit tests may influence how instructors design courses and assignments. The presence of richer, more adaptable feedback can free educators from repetitive, individualized guidance and allow them to focus on higher-level interventions, curriculum refinement, and addressing common conceptual misunderstandings observed in aggregate. For students, especially novices, having both immediate correctness checks and guidance on how to think differently about their problems can accelerate the transition from simple debugging to thoughtful, strategic problem-solving. \\

\noindent\textbf{Key Findings and Opportunities for Advancement:} 
\noindent These results open multiple avenues for future exploration. Investigating how different subgroups of students, varying in their prior experience, confidence, or learning styles, respond to AI hints versus unit test feedback could help tailor feedback mechanisms to more individual needs. Additionally, exploring long-term impacts on learning outcomes, such as whether students who engage more deeply with AI Feedback subsequently perform better in advanced courses or in exams where there is no AI Feedback, would provide valuable insights. 

In summary, these findings suggest that feedback in APASs can be improved significantly by combining unit tests for reliable checks with AI for dynamic, adaptive hints that nurture deeper engagement. Such a balanced environment holds promise for supporting students in becoming more confident, capable programmers who not only meet assignment requirements but also internalize essential problem-solving competencies. With ongoing refinements and careful consideration of trust, scalability, and infrastructure, these insights point toward more sophisticated and pedagogically effective feedback ecosystems that benefit both learners and educators.

\section{Limitations and Threats to Validity}
\label{sec:limitations}

In line with the guidelines proposed by \cite{wohlin2012experimentation}, we discuss our study's limitations across four primary categories: \emph{Construct Validity}, \emph{Internal Validity}, \emph{External Validity}, and \emph{Conclusion Validity}.

\subsection{Construct Validity} Construct validity concerns the degree to which the study accurately captures and measures the constructs it aims to investigate, in our case, the impact of different feedback forms in APASs on student performance and engagement.

\textbf{Choice of Exercises and Feedback Types.} While the exercises and test cases used were carefully peer-reviewed to align with the core competencies of introductory programming courses taught at the two universities, they may not fully encompass the breadth of difficulties, topics, or problem-solving approaches found in all programming curricula. More diverse or context-rich exercises could produce different interaction patterns, particularly with certain feedback types (e.g., AI-based hints). Similarly, our selection of feedback mechanisms may not represent the full range of feedback modalities available in all APASs, limiting the ability to generalize to other forms of guidance (e.g., video or peer feedback).

\textbf{Operationalization of ``AI-Guided Feedback``.} We used GPT-3.5-Turbo to provide on-demand suggestions and clarifications. Although this model generally produced high-quality output, the nature of AI-generated responses—ranging from concise hints to more elaborate explanations—can vary with prompt formulation and server load. Occasional ambiguity or inaccuracies in these responses could have affected students’ trust and willingness to rely on AI Feedback, influencing both engagement and outcomes.

\subsection{Internal Validity} Internal validity pertains to whether the observed effects (e.g., performance gains, time spent on tasks) are truly attributable to the different feedback forms rather than confounding factors.

\textbf{Teaching Setting and Instructional Method.} Although the experiment was designed with standardized guidelines, variations in how different instructors integrated and emphasized the APAS in their curricula could have introduced discrepancies in student motivation and preparedness. Furthermore, students received bonus points for participating in the experiment; this incentive may have triggered a self-selection bias, as students highly motivated by extra credit or more comfortable programming could be overrepresented. Despite these concerns, such incentives helped secure a sufficiently large sample size and are common practice in educational research.

\textbf{Familiarity and Comfort with the APAS.} Students entered the experiment with varying levels of prior experience using the APAS (Artemis). Those who had used it extensively may have spent less cognitive effort on interface navigation and more on solving the core programming challenges. Conversely, students with minimal exposure may have been disproportionately distracted by platform logistics. Although we required all participants to complete at least one practice exercise on Artemis before the experimental session, the depth and quality of that engagement could not be fully controlled.

\textbf{Reliability of AI Feedback.} The AI-generated feedback itself can also be viewed as a source of potential bias. Inconsistencies in the model's performance, influenced by prompt phrasing or model updates, may have affected the reliability of suggestions offered to different students at different times. This variation in guidance could alter both perceived and actual improvements in code quality or problem-solving strategy, thus confounding the internal validity of the comparison between feedback conditions.

\textbf{Timing measures and message length.}
\revmod{Longer intervals between submissions in the AI condition may reflect more content to read in AI responses rather than deeper processing. Our unit tests were defined rigorously with detailed failure messages, which should narrow any gap in reading time relative to AI outputs. However, we did not statistically analyze the length or verbosity of AI responses, nor did we normalize timing by message length. Accordingly, the timing results are correlational and should be interpreted with caution.}

\subsection{External Validity} External validity addresses the generalizability of the findings to other settings, populations, and APASs.

\textbf{Representativeness of the Student Sample.} While the experiment was conducted across multiple universities, the sample primarily consisted of students enrolled in introductory programming courses. Differences in educational backgrounds, learning styles, and personal motivations still exist within this population, but additional variety, such as professional developers or non-computer-science majors, could yield different insights.
\revmod{Moreover, only a small fraction of participants reported extensive prior programming experience, and this subgroup was too small and heterogeneous for meaningful analysis.}
Future studies involving more diverse learner groups may reveal more nuanced effects of AI-driven and traditional feedback.

\textbf{Representativeness of Artemis as an APAS.} Artemis shares many features common to modern APASs (e.g., automated grading, iterative feedback, analytics dashboards). However, variations in interface design, integration with external development tools, and customizable feedback pipelines mean that not all APASs function exactly like Artemis. Consequently, our observations, while indicative of broader trends, may not transfer perfectly to platforms with different architectures, user interfaces, or pedagogical philosophies.

\subsection{Conclusion Validity} Conclusion validity pertains to the degree of confidence we can have in the relationship between the treatments (different feedback forms) and the observed outcomes (student performance, engagement metrics).

\textbf{Measurement of Student Engagement and Performance.} Our primary metrics included time between submissions, total number of submissions, and final solution correctness. While these indicators provide valuable quantitative insights, they do not fully capture the qualitative dimensions of engagement—such as depth of understanding or the specific cognitive processes students employ. The increased time between submissions in the AI-guided group, for instance, might partly reflect reading and interpreting AI Feedback rather than deeper conceptual engagement.

\textbf{Interpretation of Observed Patterns.} Although significant differences emerged between AI-driven and traditional feedback, caution is warranted in attributing improvements solely to the type of feedback. Other unmeasured variables, such as problem difficulty or a student's prior knowledge of the specific topic, may have influenced outcomes. Future studies employing additional qualitative measures (e.g., think-aloud protocols and interviews) could help clarify the causal pathways behind these observed patterns.

\section{Conclusion \& Future Work}
\label{sec:conclusion}

In this study, we present a detailed examination of feedback mechanisms in automated programming assessment systems, focusing on compiler-based, unit test-based, and AI-driven feedback. The results highlight the importance of tailored feedback in enhancing student engagement, improving problem-solving strategies, and fostering deeper learning outcomes.

Unit test feedback was perceived as the most reliable and familiar by students, aligning closely with their expectations for correctness verification. However, it was the AI-driven feedback that led to the highest correctness scores and encouraged more iterative and thoughtful problem-solving. Compiler feedback, while effective for simple syntax and runtime errors, showed limited utility for addressing complex programming challenges, which negatively affected both student performance and engagement.

The findings suggest that combining unit test and AI Feedback can offer significant pedagogical advantages. Unit tests provide clear, objective correctness checks, while AI Feedback delivers adaptive, context-aware guidance. This combination can support students in developing not only technical accuracy but also critical thinking and more sophisticated skills. Implementing a hybrid feedback mechanism in APASs could create a more supportive learning environment, particularly for novice programmers, by addressing diverse needs and learning styles.

The results also point to key challenges and opportunities for further research. For instance, while AI Feedback demonstrated the potential for fostering deeper engagement, inconsistencies and occasional hallucinations in suggestions highlight the need for improved reliability. Integrating real-time error detection and chatbot-like interaction capabilities would enhance the usability and trustworthiness of AI-driven feedback ~\cite{frankford2024ai, bassner2024iris}. Additionally, further exploration is needed to tailor feedback to different student cohorts, such as those with varying levels of prior programming experience or confidence.

As part of our ongoing and future research, we plan to further investigate the long-term effects of combined feedback mechanisms on programming education. Evaluating how students who use such systems perform in advanced courses or real-world programming tasks could provide valuable insights into the actual impact of these feedback approaches throughout subsequent programming courses. Moreover, analyzing how different subgroups, such as students from diverse academic backgrounds, benefit from these feedback types may help refine and optimize APAS designs.
\revmod{Future work should also evaluate a more interactive, chat-based AI tutor that allows personalized follow-ups and measure its effects on engagement and performance.
Another interesting direction for future work is to investigate how more extensive prior programming experience influences students’ preferences for different feedback types. Although only a very small share of our participants reported substantial prior experience, future studies with more heterogeneous cohorts could provide deeper insights. This may also open possibilities for adapting AI tutor responses to students’ varying skill levels and backgrounds, thereby offering more tailored support.
}

From a technical perspective, enhancing the scalability and cost-efficiency of AI Feedback systems will be crucial for their broader adoption. As large language models continue to evolve, optimizing their deployment within educational contexts will require balancing performance, resource consumption, and accessibility. Collaborative efforts between academia, industry, and educational institutions will play a pivotal role in addressing these challenges.